%% file: paper.tex
\documentclass[acmsmall]{acmart}

\usepackage[normalem]{ulem}
\usepackage{xspace}
\usepackage{latexsym}
\usepackage{graphicx}
\usepackage{subcaption}
\usepackage{verbatim} 
\usepackage{multirow}
\usepackage[export]{adjustbox}

\usepackage[nameinlink]{cleveref}
\crefname{section}{\S}{\S\S}

\usepackage{wasysym}
\usepackage{pifont}
\newcommand{\cmark}{\textcolor{green}{\ding{51}}}
\newcommand{\xmark}{\ding{55}}

\usepackage{cleveref}
\crefname{section}{�}{��}
\Crefname{section}{�}{��}

\usepackage{soul}
\usepackage{booktabs}

\usepackage[mathscr]{euscript}
\usepackage{fancyhdr}
\newcommand{\etal}{\emph{et\,al.}\xspace}

\newcommand{\Changes}[1]{\textcolor{red}{#1}}

\newcommand{\hardVal}{19.0\%}

\setcopyright{rightsretained}
\begin{document}

\date{}


\title{Understanding Persistent-Memory Related Issues  in the Linux Kernel
}

\author{Om Rameshwar Gatla, Duo Zhang, Wei Xu, Mai Zheng}
\affiliation{
	\institution{\\ \textit{Department of Electrical and Computer Engineering, Iowa State University}}
	\city{}
   \country{}
}
%
 
\begin{abstract}
\input{sys_abstract}
\end{abstract}

\ccsdesc[500]{Software and its engineering~Operating systems}
\ccsdesc[300]{Hardware~Memory and dense storage}

\keywords{Persistent Memory, Kernel Patches, Bug Detection, Reliability}

\maketitle



\input{sys_intro.tex}
\input{sys_unique.tex}
\input{sys_method.tex}

\input{sys_patchoverview.tex}
\input{sys_pmbugs.tex}
\input{sys_pmtools.tex}
\input{related}

\input{sys_discussion}
\input{conclude}
\input{acknowledge.tex}

\bibliographystyle{plain}
\interlinepenalty=10000
\bibliography{reference}
\end{document}

%% file: sys_abstract.tex

Persistent memory (PM) technologies have inspired 
a wide range of  PM-based system optimizations. 
However, building correct PM-based systems is difficult due 
to the unique characteristics of PM hardware.
To better understand the challenges as well as the opportunities to address them,
this paper presents a comprehensive study of PM-related issues in the Linux kernel.
By analyzing 1,553 PM-related kernel patches in depth and conducting experiments on reproducibility and tool extension,
we derive multiple insights  in terms of PM patch categories, PM bug patterns, consequences,  fix strategies, triggering conditions, and remedy solutions.
We hope our results could contribute to the development of  robust PM-based storage systems.

%% file: sys_intro.tex
\section{Introduction}
\label{sec:intro}

Persistent memory (PM) technologies
offer attractive features for developing storage systems and applications. 
{For example, phase-change memory (PCM)~\cite{pcm}, spin-transfer torque RAM (STT-RAM)~\cite{STTRAM}, Intel's infamous Optane\textsuperscript{\texttrademark} DCPMM~\cite{IntelDCPM}, and the promising vendor-neutral CXL-based PM technologies~\cite{cxl-pmem,bhardwaj2022cache} can  support byte-granularity accesses with close to DRAM latencies, while also providing durability guarantees.}
Such new  properties have inspired 
a wide range of  PM-based software optimizations~\cite{XXXnvdimmdriivers,PMwearmanagmentpagiing,NOVA,PMFS,PMindexingXXX,TC23-DataDistribution}. 

Unfortunately, building correct PM-based software systems is challenging~\cite{lee2019recipe,XXXmanyPMworks}.
For example, to ensure persistence,  PM writes must be flushed from CPU cache explicitly via specific instructions (e.g., \texttt{clflushopt}); to ensure ordering,  memory fences  must be inserted (e.g., \texttt{mfence}). 
Moreover, to manage PM devices and support PM programming libraries 
(e.g., PMDK~\cite{pmdk}), 
 multiple OS kernel subsystems must be revised (e.g., \texttt{dax}, \texttt{libnvdimm}).
 Such complexity could potentially lead to obscure bugs that hurt system reliability and security. 
 

Addressing the challenge above will require cohesive efforts from multiple related directions including PM bug detection~\cite{AGAMOTTO-OSDI20,liu2019pmtest,liu2020cross,PersistenceInspector,pmemcheck}, PM programming support~\cite{pmdk}, PM specifications~\cite{pminterface}, among others.
All of these directions will  benefit from a better understanding of real-world PM-related bug characteristics.

Many studies have been conducted to understand and guide the improvement of software~\cite{lu2008learning,Lu-FAST13-FS,chou2001empirical,lazar2014does,chen2011linux,gunawi2014bug}.
For example, Lu \etal{}~\cite{Lu-FAST13-FS} studied 5,079 patches of 6 Linux file systems 
and derived various patterns of file system evolution; {the  study has inspired various follow-up research on  file systems reliability~\cite{Changwoo-SOSP15-CrosscheckingFS,Om-FAST18-RFSCK} and the dataset of file system bug patches has been directly used for evaluating the effectiveness of new bug detection tools \cite{Changwoo-SOSP15-CrosscheckingFS}. While influential, this study does not cover PM-related issues, as the direct-access (\texttt{dax}) feature of file systems was introduced after this study was performed. More recently, researchers have studied PM related bug cases. For example,}
Neal \etal~\cite{AGAMOTTO-OSDI20} studied  63 PM bugs (mostly from {the PMDK library}~\cite{pmdk})  and 
identified two general patterns of PM misuse.
While these existing efforts have generated valuable insights for their targets,
they do not cover the potential PM-related issues in the Linux kernel. 

In this work, we perform the first comprehensive study on PM-related bugs in the Linux kernel. 
We focus on the Linux kernel for its prime importance in supporting PM programming~\cite{dax,libnvdimm,nfit}.
Our study is based on 1,553 PM-related patches committed in Linux between Jan. 2011 and Dec. 2021, spanning over 10 years. 
For each patch, we  carefully examine its purpose and logic, 
which enables us to gain quantitative insights along multiple dimensions:

 {First, we observe that a large number of PM patches (38.9\%) are for maintenance purpose, and a similar portion (38.4\%) are for adding new features or improving the efficiency of existing ones. These two major categories  reflect the significant efforts needed to add PM devices to the Linux ecosystem and to keep the kernel well-maintained.}{ Meanwhile, a non-negligible portion (22.7\%) are bug patches for fixing correctness issues.}

Next,  we analyze the PM bug patches in depth. 
We find that the majority of  kernel subsystems have been involved in the bug patches (e.g., `\texttt{arch}', `\texttt{fs}', `\texttt{drivers}', `\texttt{block}', `\texttt{mm}'),  
with  drivers  and file systems being the most ``buggy'' ones. 
This reflects the complexity of implementing the PM support correctly in the kernel, especially the \texttt{nvdimm} driver and the \texttt{dax} file system support.

In terms of bug patterns, we find that the classic semantic and concurrency bugs remain pervasive in our dataset (49.7\% and 14.8\% respectively), although the root causes are different.
Also, many PM bugs are uniquely dependent on hardware (19.0\%), which may be caused by misunderstanding of 
specifications, miscalculation of addresses, etc.
Such bugs may lead to missing devices, inaccessible devices, or even security issues, among others.

In terms of bug fixes, we find that PM bugs tend to require more lines of code to fix compared to non-PM bugs reported in previous studies~\cite{Lu-FAST13-FS}. {Also, 20.8\% bugs  require modifying multiple kernel subsystems to fix}, 
which implies the complexity. 
In the extreme cases (0.9\%),  developers may {temporarily} ``fix'' a PM bug by disabling a PM feature, hoping for a major re-work in the future.
On the other hand, we  observe that  different PM bugs may be fixed in a similar way by refining the sanity checks. 

Moreover,  to better understand the  conditions for manifesting the issues and help develop effective remedy solutions,  we identify a subset of bug patches with relatively complete information, and 
attempt to reproduce them experimentally. 
We find that configuration parameters in different utilities (e.g., \texttt{mkfs} for creating a file system, \texttt{ndctl} for managing the \texttt{libnvdimm}  subsystem) are critically important for manifesting the issues, which suggests that it is necessary to take into account the configuration states when building bug detection tools.

Finally, we look into the potential solutions to address the PM-related issues in our study. We examine multiple representative PM bug detectors~\cite{AGAMOTTO-OSDI20,liu2020cross,liu2019pmtest}
and find that they are largely inadequate for addressing the PM bugs in our study. 
 On the other hand, a few recently proposed non-PM bug detectors~\cite{krace-ieee-sp20,razzer-ieee-sp19,janus-ieee-sp19,hydra-sosp19} could potentially be applied to detect a great portion of PM bugs if a few  common challenges (e.g., precise PM emulation, PM-specific configuration and workload support) are addressed. To better understand the feasibility of extending existing tools for PM bug detection, we further extend one existing bug detector called Dr.Checker~\cite{drchecker} to analyze  PM kernel modules. By adding PM-specific modifications, the extended Dr.Checker, which we call Dr.Checker+, can successfully analyze the major PM code paths in the Linux kernel. While the effectiveness is still  limited by the capability of the vanilla Dr.Checker, we believe that extending existing tools to make them work for the PM subsystem  can be an important   first step towards  addressing the PM-related challenges exposed in our study.
 
 Note that this manuscript is extended from a conference version~\cite{pm-bug-study}. The major changes include: (1) collecting and analyzing  one new year of PM-related patches (i.e., Jan. to Dec. 2021), (2) conducting reproducibility experiments and identifying manifestation conditions,  (3) analyzing more existing tools and extending Dr.Checker for analyzing PM driver modules, (4) adding background, bug examples, etc. to make the paper more  clear and complete. We have released our study results including the  dataset and the extended Dr.Checker+ publicly on Git~\cite{dsl-pmbug-repo}. We hope our study could contribute to the development of effective PM bug detectors and the enhancement of robust PM-based systems.

{The rest of the  paper is organized as follows: 
\S\ref{sec:unique} provides background on the unique hardware and software characteristics of PM devices;
\S\ref{sec:methodology} describes the study methodology; 
\S\ref{sec:overview} presents the overview of PM patches; 
  \S\ref{sec:pmbugs} characterizes PM bugs in details; 
  \S\ref{sec:reproduce} presents our experiments on PM bug reproducibility;
  \S\ref{sec:implications} discusses the implications on bug detection including our extension to  Dr. Checker for analyzing PM subsystem in the Linux kernel;
  \S\ref{sec:related}  discusses related work and \S\ref{sec:conclude} concludes the paper.}

%% file: sys_unique.tex
\section{Background}
\label{sec:unique}

\begin{figure}
\subfloat[PM Persistence Domains ]{\includegraphics[width = 2in]{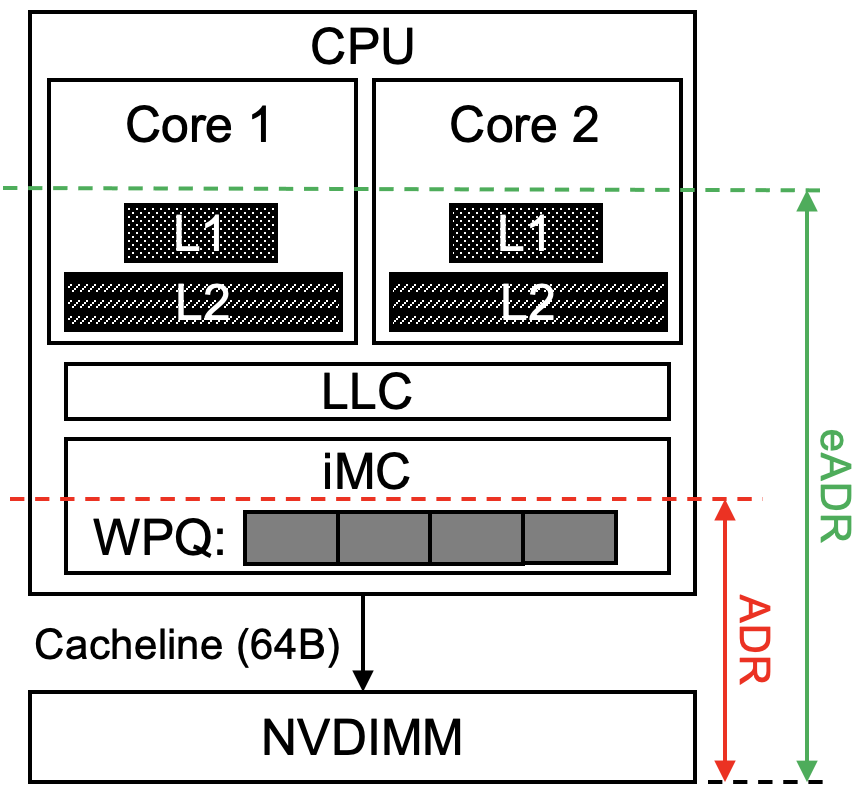}} \hspace{.5cm}
\subfloat[Software Abstractions of NVDIMM Subsystem in Linux]{\includegraphics[width = 3in]{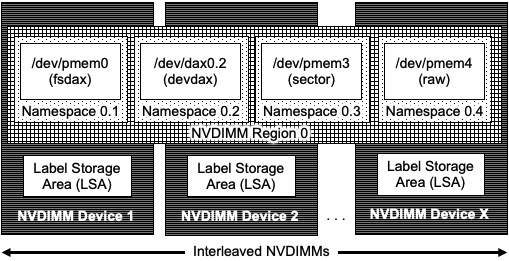}}
\caption{Background of PM Hardware and Software}
\label{fig:PM_feature}
\end{figure}

In this section, we introduce the characteristics of PM hardware and software which may contribute to correctness issues in PM-based systems.

\subsection{NVM and PM Device Types}
{
A wide range of non-volatile memory (NVM) devices have been proposed and are being developed with different degrees of maturity. To facilitate standardization, the JEDEC specification \cite{jedec-website} classifies NVM devices into three types based on the method used for persistence:}

\begin{itemize}
    \item 
{
NVDIMM-N devices employ both DRAM and Flash modules, and is the initial form of PM devices developed. These devices achieve non-volatility by copying contents in the DRAM to the flash modules when the host power is lost using an energy source managed by either the device (e.g., on-chip capacitor) or the host. AGIGA's NVDIMM-N \cite{agiga-nvdimm} and HPE NVDIMM \cite{hpe-nvdimm} devices are a few examples.
}

\item
{
NVDIMM-F devices use NAND flash modules over the memory bus as storage medium. Some eamples of this type are SanDisk's ULLtraDIMM \cite{sandisk-ulltradimm} and IBM's exFlash DIMM \cite{ibm-exflash}. However, due to their high access latency, these devices have largely been discontinued.
}

\item
{
NVDIMM-P standard encompasses devices that employ new storage technologies such as  PCM~\cite{pcm}, STT-RAM~\cite{STTRAM}, ReRAM~\cite{chua1971memristor}, etc. to achieve persistence. The commercialized Intel\textsuperscript{\textregistered} Optane\textsuperscript{\texttrademark} DCPMM and the vendor-neutral CXL-based PMs also belong to this category. 
}

\end{itemize}

{
In addition, to facilitate NVM programming, the Storage Networking Industry Association (SNIA~\cite{SNIA}) differentiates the concepts of NVM and PM~\cite{SNIA-NVM-ProgrammingGuide}:
NVM refers to any type of memory-based, persistent media (including flash memory packaged as solid state
disks), while PM refers to a subset of NVM  technology with performance characteristics suitable for a load and store programming (e.g., PCM, STT-RAM, Optane DCPMM). 
In this paper, we follow the definition and study the Linux kernel issues related to such PM devices.}

\subsection{PM Device Characteristics}

PM devices reside on the memory bus along with DRAM devices. 
Unlike traditional block storage devices that provide a block IO interface to access data, PM devices 
use memory load and store instructions. 
{
These devices typically offer lower access latency and higher endurance compared to flash-based storage devices. For example, STT-RAM and PCM devices may offer latencies in the range of 5 - 220 nanoseconds with an endurance rate around $10^{12}$ writes per bit \cite{yang2015nv}.}

{
Among existing PM devices, Intel\textsuperscript{\textregistered} Optane\textsuperscript{\texttrademark} DCPMM has been the most prominent one.  
Great efforts have been made to integrate the device to the existing ecosystem. For example, new programming model with special CPU instructions (e.g., \texttt{clwb}, \texttt{clflush} and \texttt{clflushopt}) were introduced to provide durability guarantees. In addition, new persistence domains were defined to better understand the persistence guarantees offered by these devices. 
Fig. \ref{fig:PM_feature}(a) shows the two persistence domain supported on Intel platforms. {Asynchronous DRAM Refresh (ADR)} domain specifies that all updates that reach the write pending queue (WPQ) in the integrated memory controller (iMC) are guaranteed to be persisted in an event of system failure. Whereas, {Enhanced ADR (eADR)} domain specifies that all updates in the caches and memory controller are persisted in an event of system failure. Therefore, there is no need for developers to explicitly flush cachelines using special CPU instructions. Note that eADR is a relatively recent feature which may not be available in all platforms. Also, while eADR may provide stronger persistence guarantee at the hardware level, PM software still needs to be carefully designed to achieve desired high level properties and correctness guarantees (e.g., atomicity of a transaction) \cite{bhardwaj2022cache}. These efforts have inspired the development of many new PM-optimized software \cite{pmdk,hennecke2020daos}.
}

{Most recently, vendor-neutral  interfaces such as Compute Express Link (CXL) have been pushing the evolution of PM devices further. For example, CXL 2.0 specification has included the support for PM devices~\cite{cxl,cxl-pmem}. These CXL-based PM devices are expected to be compatible with the existing NVDIMM specifications and they will still show up as special memory devices to the operating system (OS) kernel.
Therefore, although Intel is winding down the Optane DCPMM business with limited support in the next  3 to 5 years~\cite{intel-pmemio-update}, the community is expecting to see CXL-based PM devices in the near future which are  compatible with the  NVDIMM ecosystem~\cite{cxl,cxl-pmem,bhardwaj2022cache,intel-pmemio-update}.
}
\subsection{PM Software Abstractions}
To support PM devices in the storage system, multiple subsystems in the Linux kernel were modified. New Unified Extensible Firmware Interface (UEFI)  specifications were introduced and new NVDIMM Firmware Interface Table (NFIT) driver was added to the OS kernel~\cite{nfit}. 
  Moreover, to manage and access  PM devices efficiently,
the Linux kernel exposes the PM storage space over multiple abstractions.
 Specifically, the PM devices attached to NVDIMM can be configured into either interleaved  or  non-interleaved mode.  
In the interleaved mode (Figure \ref{fig:PM_feature}b),  an NVDIMM region is created over multiple NVDIMM devices; in the non-interleaved mode, each device has one NVDIMM region. Additionally, multiple namespaces may be created within one NVDIMM region, which is conceptually similar to creating multiple partitions on one block device.



The NVDIMM devices are registered as special memory devices in the Linux kernel.
Each NVDIMM namespace maintains a dedicated memory area to store Page Frame Number (PFN) metadata, which supports the necessary address translation when accessing the storage media.
%
%
Moreover, to bypass the page cache and enable direct access to PM, major file systems such as Ext4 and XFS have introduced the direct-access (DAX) feature based on the traditional Direct IO feature. As will be shown in our study, however, it is challenging to implement such PM-oriented optimizations correctly due to the system complexity.  



%% file: sys_method.tex
\section{Methodology}
\label{sec:methodology}

In this section, we  describe how we collect 
the dataset for study (\S\ref{sec:collection}), 
how we characterize the PM-related patches and bugs (\S\ref{sec:characterizationmethods}),
and the limitations of the methodology (\S\ref{sec:limitations}).
\subsection{Dataset Collection and Refinement}
\label{sec:collection}
All changes to the Linux kernel occur in the form of patches~\cite{kernelpatch}, including but not limited to bug fixes.
We collect  PM-related  patches from the Linux source tree for study via four steps as follows:

First, we collect all patches committed to the Linux kernel between Jan. 2011 and Dec. 2021, 
which generates a dataset containing {about 772,000 patches}. 

Second, in order to effectively identify PM-related patches, we refine the dataset using a wide set of PM-related keywords, such as `\texttt{persistent memory}', `\texttt{pmem}',  
`\texttt{dax}', `\texttt{ndctl}', `\texttt{nvdimm}', 
`\texttt{clflushopt}', etc. 
The resulting dataset contains {3,050} patches. 
Note that this step is similar to the keyword search in  previous studies~\cite{lu2008learning,gunawi2016does}.

Third, to prune the potential noise,  we refine the dataset further by manual examination. 
Each patch is analyzed at least twice by different researchers, and those irrelevant to PM are excluded based on our domain knowledge. 
{The final dataset 
contains 1,553 PM-related patches in total.}




\subsection{Dataset Analysis and Experiments}
\label{sec:characterizationmethods}
\begin{table}[tb]
    \small
	\begin{center}
        \begin{tabular}{ c | l | c }
			\textbf{Category} & \hspace{2.7cm}\textbf{Description} & \textbf{Overall}\\
			\hline
 {Bug} & Fix existing correctness issues & 352 \\
       & (e.g., misalignment of PM regions, race on PM pages) & (22.7\%)\\
			\hline
      {Feature} & Add new features or improve efficiency of existing ones & 597 \\
      	& (e.g., extend device flags, reduce write overhead) & (38.4\%)\\
			\hline
      {Maintenance } & Polish source code, compilation scripts, and documentation & 604\\
      	& (e.g., delete obsolete code, fix compilation errors) & (38.9\%)\\
      	\hline
      	\hline
			\multicolumn{2}{r|}{\textit{Total}} & 1553\\
		\end{tabular}
    \end{center}
    \caption{{Three Categories of PM-related Patches and the Percentages.}}
	\label{tab:classify}
	\vspace{-0.2in}
\end{table}

{Based on the 1,553 PM-related patches}, 
we conduct a comprehensive study
to answer four set of questions:

\begin{itemize}
    \item Overall Characteristics: What are the purposes of the PM-related patches? How many of them are merged to fix correctness issues (i.e., PM bugs)? 

    \item Bug Characteristics: What types of PM-related bugs existed in the Linux kernel? 
    What are the bug patterns and consequences? How are they fixed? 

    \item Reproducibility: Can the PM-related bugs be reproduced for future research? What are the critical conditions for manifesting the issues? 
    
    \item Implications: What are the limitations of existing PM bug detection tools?
    What are the  opportunities?
    
\end{itemize}

To answer these questions, we manually analyzed each patch in depth to understand its purpose and logic.
The patches typically follow a standard format containing a description and code changes~\cite{kernelpatch}, which enables us to characterize them along multiple
dimensions. 
For patches that contain limited information, we further looked into relevant source code
and design documents. Moreover, we conduct experiments to validate the reproducibility of PM bugs and the capability of state-of-the-art bug detectors. We present our findings for the four sets of questions above in  \S\ref{sec:overview}, \S\ref{sec:pmbugs},  \S\ref{sec:reproduce}, and \S\ref{sec:implications}, respectively.

\subsection{Limitations}
\label{sec:limitations}
The results of our study should be interpreted with 
the method in mind.
The  dataset was refined via 
PM-related keywords and manual examination,
which might be incomplete.
Also, 
we only studied PM bugs that have been triggered and
fixed in the mainline Linux kernel, 
which is biased: 
there might be other latent (potentially trickier) bugs  not yet discovered.
Nevertheless, we believe our study is one important step toward addressing the challenge.
We release our results publicly to facilitate follow-up research~\cite{dsl-pmbug-repo}.

%% file: sys_patchoverview.tex
\section{PM Patch Overview}
\label{sec:overview}
\begin{figure}[tb]
	\centering
    \includegraphics[width=3.5in]{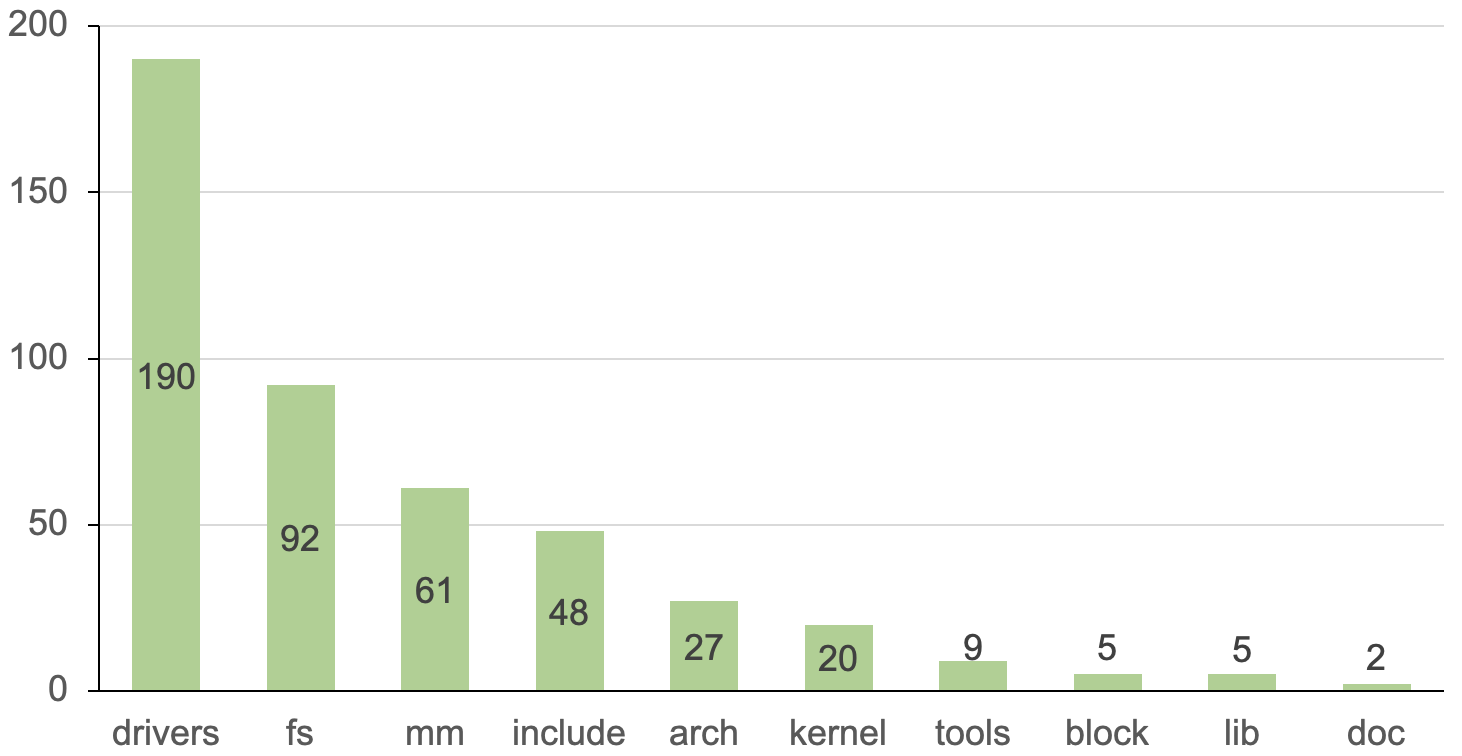}
    \vspace{-0.1in}
	\caption{ {{\bf Counts of PM Bug Patches in the Kernel Source Tree}. The figure shows that PM bug patches involve many major kernel subsystems, with drivers and file systems  contributing the most.}}
	\label{fig:folders_num_files2}
    \vspace{-0.2in}
\end{figure}




We classify all PM-related patches into three categories as shown in Table~\ref{tab:classify}:
(1) `\texttt{Bug}' means fixing existing correctness issues (e.g.,  misalignment of NVDIMM namespaces); 
(2) `\texttt{Feature}' means adding new features (e.g., extend device flags) or improving the efficiency of existing designs;
(3) `\texttt{Maintenance}' means code refactoring, compilation or documentation update, etc. 

Overall, the largest category in our dataset is `\texttt{Maintenance}' (38.9\%), 
which is consistent with previous studies on Linux file system patches~\cite{Lu-FAST13-FS}. This reflects the significant effort needed to keep PM-related kernel components well-maintained. We find that the majority of maintenance patches are related to code refactoring (e.g., removing deprecated functions or driver entry points), and occasionally, the refactoring code may introduce new correctness issues that need to be fixed via other patches.

The second largest category is `\texttt{Feature}' (38.4\%). 
This reflects the significant changes needed to add PM to the Linux ecosystem 
which has been optimized for non-PM devices for decades.
One interesting observation is that many (40+) feature patches are proactive 
(e.g., \textit{``In preparation for adding more flags, convert the existing flag to a bit-flag''}~\cite{A-Proactive-Patch}),
which  
may imply that PM-based extensions tend to be well-planned in advance.
Also,  
most recent feature patches are related to supporting PM devices on the  Compute eXpress Link (CXL) interface~\cite{cxl},  which indicates the rapid evolution of PM relevant techniques. 

 
The `\texttt{Bug}' patches, which directly represent confirmed and resolved correctness issues in the kernel,  account for a non-negligible portion (i.e., 22.7\% overall).
 We analyze this important set of patches further 
in the next section.    

%% file: sys_pmbugs.tex
\section{PM Bug Characteristics}
\label{sec:pmbugs}

\input{sys_buglocation.tex}

\subsection{Bug Pattern}
\label{sec:bugtype}

\begin{table}[tb]
	\scriptsize
	\begin{center}
		\begin{tabular}{ c | r | l | l }
			\textbf{Type} & \textbf{Subtype}\hspace{1.5mm} & \hspace{12.5mm}\textbf{Description} & \textbf{Major}\\
			& & & \textbf{Subsystems} \\
			\hline
            \multirow{7}{*}{\rotatebox{90}{\parbox{1.3cm}{\textbf{Hardware} \textbf{Dependent}}}}          & Specification & misunderstand specification & drivers, arch, \\
           		& &(e.g.: ambiguous ACPI specifications) & include\\
			\cline{2-4}
			& Alignment & mismatch b/w abstractions of PM device & drivers, mm, \\
			& & (e.g.: misaligned NVDIMM namespace) & arch \\
			\cline{2-4}
            		& Compatibility & Device or architecture compatibility issue & drivers, arch, mm\\
			\cline{2-4}
			& Cache & Misuse of cache  related operations & arch, mm, \\
			&    & (e.g.: miss cacheline flush)& drivers \\
			\hline
			\multirow{4}{*}{\rotatebox[origin=c]{90}{\textbf{Semantic}\hspace{2mm}}} & Logic & improper design & drivers, fs, mm\\
			& & (e.g.: wrong design for DAX PMD mgmt.) & \\
			\cline{2-4}
			& State & incorrect update to PM State & fs, mm \\
			\cline{2-4}
			& Others & other semantic issues & drivers, fs\\
			&	       & (e.g.: wrong function / variable names) &\\
			\hline
			\multirow{7}{*}{\rotatebox{90}{\textbf{Concurrency}}} & Race & data race issues involving DAX IO & fs, mm, drivers\\
			\cline{2-4}
			& Deadlock & deadlock on accessing PM resource & drivers, mm, fs \\
			\cline{2-4}
			& Atomicity & violation of atomic property for PM access & drivers, fs, mm\\
			\cline{2-4}
			& Wrong Lock & use wrong lock for PM access & fs, drivers, block\\
			\cline{2-4}
			& Order & violation of order of multiple PM accesses & fs\\
			\cline{2-4}
			& Double Unlock & unlock twice for PM resource & drivers\\
			\cline{2-4}
			& Miss Unlock & forget to unlock PM resource & drivers\\
			\hline
			\multirow{4}{*}{\rotatebox[origin=c]{90}{\textbf{Memory}}} & Null Pointer & dereference null PM / DRAM pointer & drivers, fs, mm\\ 
			\cline{2-4}
			& Resource Leak & PM / DRAM resource not released & drivers, mm, arch \\
			\cline{2-4}
			& Uninit. Read & read uninitialized PM / DRAM variables & drivers, fs\\
			\cline{2-4}
			& Overflow & overrun the boundary of PM/DRAM struct. & drivers, fs, include\\
			\hline
			\multirow{3}{*}{\rotatebox[origin=c]{90}{\shortstack[l]{\textbf{Error  }\\ \textbf{Code  }}}} & Error Return & no / wrong error code returned & drivers, fs, kernel\\
			\cline{2-4}
			& Error Check & miss / wrong error check &  drivers, fs, mm\\
			& & &\\
		\end{tabular}
	\end{center}
	\caption{ {\bf Classification of PM Bug Patterns}. \textit{The last column shows
	the major subsystems (up to 3) affected by the bugs.}}
	\label{tab:bug-type}
\end{table}

\begin{figure}[tb]
	\centering
	\includegraphics[width=4in]{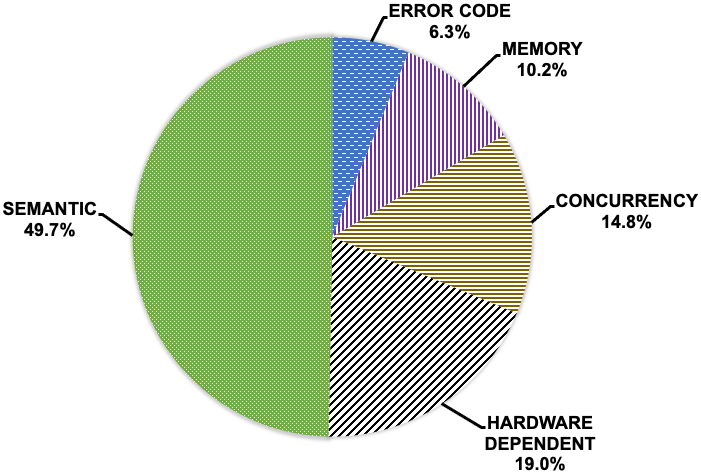}
	\vspace{-0.15in}
	\caption{ {\bf Percentages of PM Bug Types
 } 
 }
	\label{fig:bug-types}
\end{figure}
\begin{figure}[tb]
	\centering
	\includegraphics[width=5in]{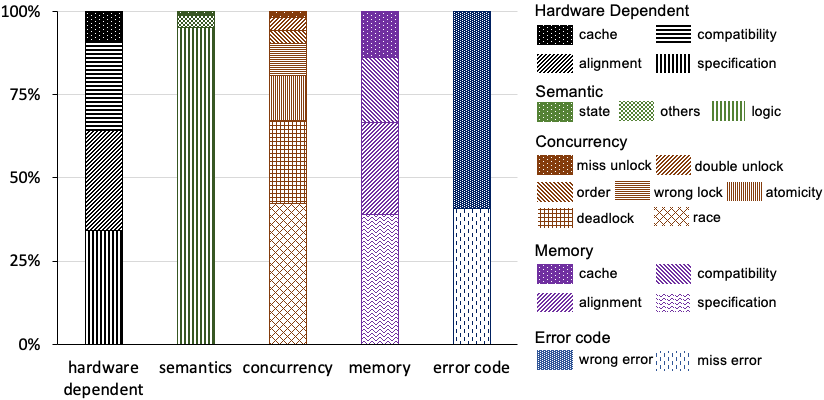}
	\vspace{-0.15in}
	\caption{ {\bf Percentages of  PM Bug Subtypes}. The five bars represent the five main types, each of which consists of multiple subtypes.}
	\vspace{-0.2in}
    \label{fig:bug-subtypes}
\end{figure}

To build  reliable and secure PM systems, it is important to understand the types of bugs occurred.  
We analyze the PM bug patches in depth and classify the bugs into five main types (Table~\ref{tab:bug-type}): 
\textit{Hardware Dependent},   \textit{Semantic}, \textit{Concurrency}, \textit{Memory} and \textit{Error Code}.
Each type includes multiple subtypes. 
The last column of Table~\ref{tab:bug-type}
shows the major subsystems   
affected by each type of bugs. 
For clarity, the column only lists up to 3 subsystems for each type.
We can see that
the same type of bugs may affect multiple subsystems (e.g., `\texttt{drivers}', `\texttt{fs}', `\texttt{mm}'),
and each subsystem may suffer from multiple types of bugs.

Figure~\ref{fig:bug-types} shows the relative percentages of the five main types. We can see that the \textit{Semantic} type is dominant (49.7\%), followed by the \textit{Hardware Dependent} type (\hardVal).
Similarly, Figure~\ref{fig:bug-subtypes} further shows the relative percentages of the subtypes within each main type.
We elaborate on multiple representative types below based on these classifications, with an emphasis on the ones that are different from previous studies~\cite{lu2008learning,Lu-FAST13-FS}.  


\subsubsection{\bf Hardware Dependent}
\label{sec:hwdependent}

Compared to previous studies~\cite{lu2008learning,Lu-FAST13-FS},
the most unique pattern observed in our dataset is \textit{Hardware Dependent}, 
which accounts for \hardVal  ~of  PM bugs (Figure~\ref{fig:bug-types}).
There are four subtypes including \textit{Specification} (34.3\% of \textit{Hardware Dependent}),   \textit{Alignment} (29.9\%),
\textit{Compatibility} (26.9\%),
and \textit{Cache} (8.9\%), which reflects four different  aspects of challenge for integrating  PM devices correctly to the Linux kernel.

\begin{figure}
    \centering
    \begin{subfigure}[b]{3.5in}
        \centering
        \includegraphics[width=\textwidth]{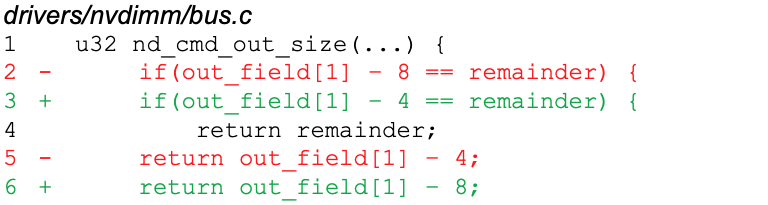}
        \caption{{\bf A Specification Bug Example}}
        \label{subfig:spec-bug}
    \end{subfigure}
    \hfill
    \begin{subfigure}[b]{3.5in}
        \centering
        \includegraphics[width=\textwidth]{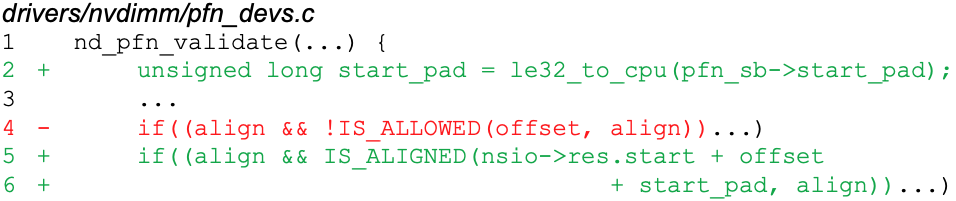}
        \caption{{\bf An Alignment Bug Example}}
        \label{subfig:alignment-bug}
    \end{subfigure}
    \vspace{0.1in}
    
    \begin{subfigure}[b]{3.5in}
        \centering
        \includegraphics[width=\textwidth]{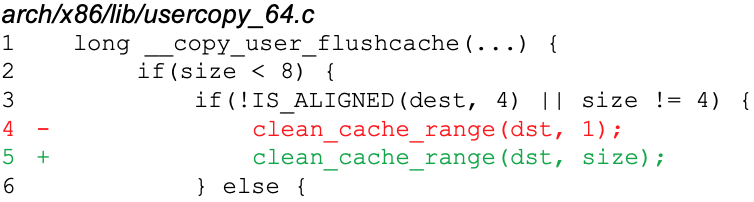}
        \caption{{\bf A Cache Bug Example}}
        \label{subfig:cache-bug}
    \end{subfigure}
    \vspace{0.1in}
    
    \caption{{Three Examples of Hardware Dependent Bugs. (a) was caused by the ambiguity of the ACPI specification; (b) was caused by misalignment between the PFN abstraction and the NVDIMM namespace; (c) was caused by inaccurate flushing of cachelines. }}
    \label{fig:pm-bugs}
\end{figure}

\textit{Specification} is the largest subtype of  \textit{Hardware Dependent} bugs (34.3\%).
Figure~\ref{subfig:spec-bug} shows an example  caused by the ambiguity of PM hardware specification.
In this case, 
the PM device uses  Address Range Scrubbing (ARS) \cite{acpi61-spec}  to communicate  errors to the kernel. ACPI 6.1 specification \cite{acpi61-spec} requires defining the size of the output buffer, but it is ambiguous if the size should include the 4-byte ARS status or not.
As a result, when the \texttt{nvdimm} driver 
should have been checking for  `\texttt{out\_field[1] - 4}', it was using `\texttt{out\_field[1] - 8}' instead, which may lead to a crash.

In terms of the other 3 subtypes,
we find that 
\textit{Alignment} issues are typically caused by 
the inconsistency between various abstractions of PM devices
(e.g., PM regions, namespaces). {Figure \ref{subfig:alignment-bug} shows an issue due to a misalignment between the Page Frame Number (PFN) device abstraction and the NVDIMM namespace. While validating the PFN device alignment, the method does not take into account of the padding at the beginning of the namespace.} This inconsistency makes the NVDIMM device inaccessible.

\textit{Compatibility} issues often arise when  the new \texttt{dax} functionality conflicts with the underlying CPU architecture or PM device.
For example, the \texttt{dax} support requires modifying page protection bits in the page table entries (PTEs) which depends on  CPU architectures. {However, PowerPC architecture does not allow for such modification to valid PTEs. A check within the memory management subsystem triggers a kernel warning in this scenario. This bug was fixed by invoking a PowerPC-specific routine when modifying PTEs.}

\textit{Cache} issues  are caused by misuse of cache-related operations (e.g., \texttt{clflushopt}). The cache-related operations have been the major focus of existing  studies on user-level PM software~\cite{AGAMOTTO-OSDI20,lee2019recipe}. Nevertheless, we find that the \textit{Cache}  subtype only accounts for 8.9\% of \textit{Hardware Dependent} bugs in our dataset (Figure~\ref{fig:bug-subtypes}).  
{Moreover, we find that the \textit{Cache} bug 
pattern is different from the typical pattern widely studied in the literature. }
Existing studies mostly focus on the cases where a specific operation (cacheline flush or memory fence) is missing. In our dataset, however, the issues may arise due to partially flushing data from the volatile CPU cache to PM media. Figure \ref{subfig:cache-bug} shows an example. In this example, the function \texttt{clean\_cache\_range} in line 4 is expected to flush the cacheline holding the data of variable \texttt{dst}. However, if the data occupy two cachelines, then the code in line 4 only flushes data in the first cacheline. A crash at this point may lead to incomplete data on persistent media. This bug was fixed (line 5) by providing the size of data as input to \texttt{clean\_cache\_range}. 
Such subtle granularity issues will likely require additional innovations on bug detection to handle.

\begin{figure}[tb]
    \centering
    \includegraphics[width=\textwidth]{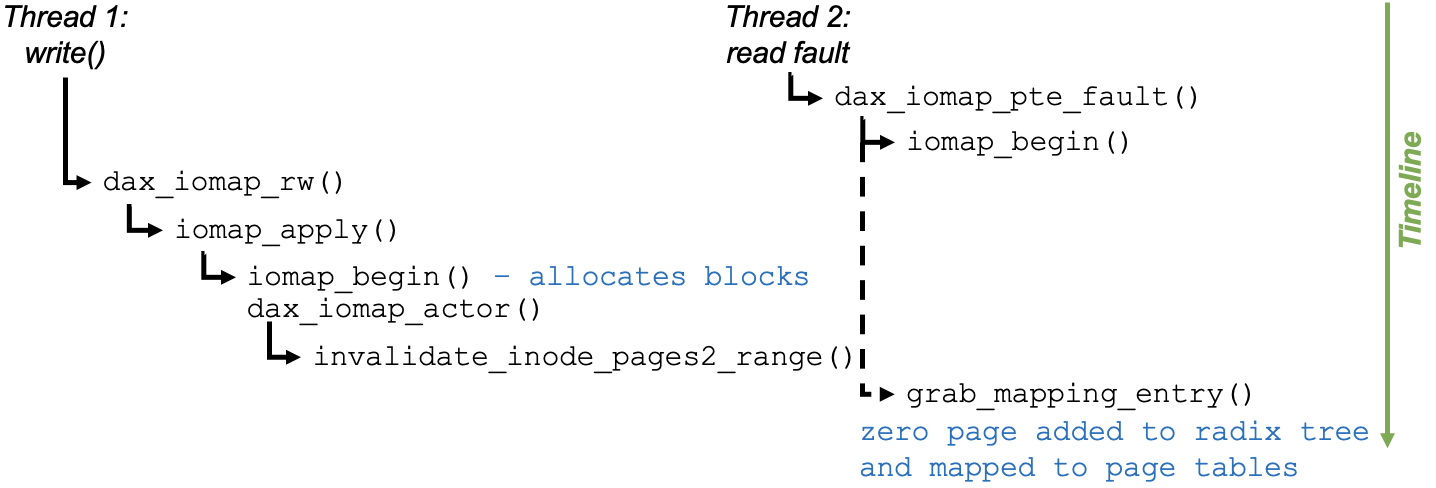}
    \caption{{A Concurrency Bug Example in the PM Context. This bug was caused
by a race condition involving DAX IO operations: Thread 1 attempts to write data to a given address range, while Thread 2 attempts to read data from the same address range.}}
    \label{fig:concurrency-bug}
\end{figure}

\begin{figure}[tb]
\centering
    \begin{subfigure}[t]{0.55\textwidth}
        \centering
        \includegraphics[width=\textwidth]{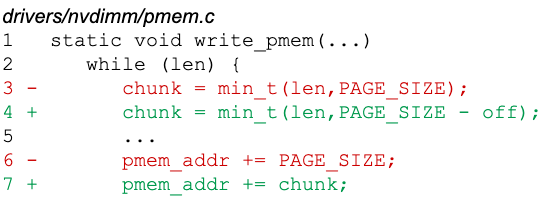}
        \caption{{\bf A Memory Overflow Bug Example}}
        \label{subfig:semantic-state-bug}
    \end{subfigure}
    ~
    \begin{subfigure}[t]{0.45\textwidth}
        \centering
        \includegraphics[width=\textwidth]{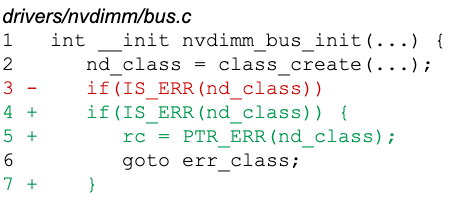}
        \caption{{\bf An Error Code Bug Example}}
        \label{subfig:error-check-bug}
    \end{subfigure}
    \begin{subfigure}[t]{\textwidth}
        \centering
        \includegraphics[width=0.8\textwidth]{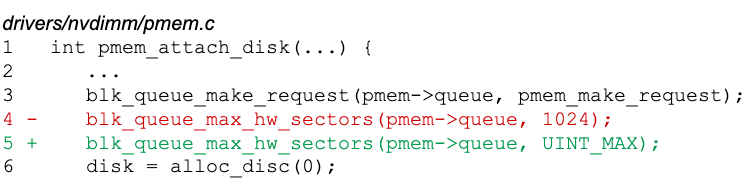}
        \caption{{\bf A Semantic Bug Example}}
        \label{subfig:semantic-logic-bug}
    \end{subfigure}
    \vspace{-0.1in}
    \caption{{Three Different Types of Classic Bug Patterns in the PM Context. (a) Memory Overflow: 
    an offset was missed when calculating a PM address  in the NVDIMM driver, which caused an out-of-bound access error;
    (b) Error Code: the return code is not captured when the struct \texttt{nd\_class} is invalid, which affected the error handling; (c) Semantic: 
    the max  size  of an I/O request for the PM device  is wrong.}}
    \label{fig:3newexamples}
\end{figure}

\subsubsection{\bf Other Types}
\label{sec:othertypes}

In addition to  \textit{Hardware Dependent}, we  find that PM  bugs may follow the classic \textit{Semantic},   \textit{Concurrency},  
 \textit{Memory}, \textit{Error Code} patterns {observations in traditional file systems and/or user-level applications~\cite{Lu-FAST13-FS,lu2008learning}, but the root causes may be different due to the different contexts}.
For example, \textit{Concurrency} bugs in our dataset are specific to the PM environment. 
 In particular, 
 we find that 
 the majority of \textit{Concurrency} PM bugs are caused by race conditions between the \texttt{dax} page fault handler and regular IO operations.
Figure~\ref{fig:concurrency-bug} shows one specific example involving two threads.
In this case,
 Thread 1 invokes a write syscall which allocates blocks on PM, and Thread 2 invokes a read to the same PM blocks which triggers a page fault.
When Thread 1 is updating the block mappings,
Thread 2 should wait until the update completes. 
However, 
due to the lack of proper locking,
Thread 2 instead maps hole pages to the page table, which results in reading zeros 
{instead of data written in place. This may cause crash-consistency issues.}
 The bug was fixed by locking the exception entry before mapping the blocks for a page fault. In this way, 
 either the writer will be blocked until read finishes or the reader will see the correct PM blocks updated by the writer.

{Figure~\ref{fig:3newexamples} shows  three additional examples of classic bug patterns in the PM context. Specifically, Figure~\ref{subfig:semantic-state-bug} shows a \textit{Memory Overflow} bug 
where an offset was missed when calculating a PM address in the NVDIMM driver, and this miscalculation led to an out-of-bound access error
{, resulting in a kernel crash.} 
In Figure~\ref{subfig:error-check-bug} (\textit{Error Code} bug), the return code was not captured when the struct \texttt{nd\_class} is invalid, which affected the error handling later 
{(i.e., no error code returned)}.
In Figure~\ref{subfig:semantic-logic-bug} (\textit{Semantic} bug), a wrong size is used for the request for accessing the NVDIMM block device, which limited the max I/O size to PM to 1024 sectors 
{resulting in an inconsistent read or write}. Overall, these classic bug patterns suggest that ``history repeats itself'', and more efforts are needed to address the classic issues in the PM context.}

\input{sys_bugconseq.tex}

\input{sys_bugfix.tex}

\input{sys_reproduce.tex}

%% file: sys_buglocation.tex
\subsection{Where Are the Bugs}
\label{sec:where}


Figure \ref{fig:folders_num_files2} shows the distribution of PM bug patches in the Linux kernel source tree. 
For clarity, we only show the major top-level directories in Linux,
which represent major subsystems (e.g., `\texttt{fs}' for file systems, `\texttt{mm}' for memory management).
In case a patch modifies multiple files across different directories (which is not uncommon as will be discussed in \S\ref{sec:bugsize}), we count it towards all directories involved. Therefore, the total count is larger than the number of PM bug patches.

We can see that `\texttt{driver}' is involved in most patches 
, which is consistent with previous studies~\cite{chou2001empirical}. {In the PM context, this is largely due to the complexity of adding the \texttt{nvdimm} driver and the support for the CXL interface.}
Also, `\texttt{fs}' accounts for the second most patches, largely due to the complexity of 
adding \texttt{dax} support for file systems~\cite{Ext4_DAX}.
The fact that  PM bug patches involve many major kernel subsystems implies that we cannot only focus on one (e.g., `\texttt{fs}') to address the challenge.

\begin{table}[tb]
	\small
	\begin{center}
        \begin{tabular}{l | c | c }
			\textbf{File Name} & \textbf{\# of }  & \textbf{LoC Changed}\\
			 & \textbf{Occurrence} & \textbf{per 100 LoC}\\
			\hline
			fs/dax.c & 41 & 5.08 \\
			\hline
			drivers/nvdimm/pfn\_devs.c & 22 & 2.62 \\
			\hline
			drivers/nvdimm/bus.c & 21 & 1.8\\
			\hline
			drivers/nvdimm/pmem.c  & 18 & 3.23\\
			\hline
			drivers/acpi/nfit/core.c & 16 & 0.7\\
			\hline
			drivers/nvdimm/region\_devs.c & 15 & 1.95\\
			\hline
			drivers/nvdimm/namespace\_devs.c & 15 & 0.8\\
			\hline
			drivers/dax/super.c & 14 & 3.27\\
			\hline
			mm/memory.c & 13 & 0.53\\
			\hline
			drivers/nvdimm/btt.c & 12 & 2.44 \\
		\end{tabular}
    \end{center}
	\caption{ {\bf Top 10 Most ``Buggy'' Files}. The table shows that 9 out of the 10 files containing the most PM bugs are  related to the `dax' or `nvdimm' supports in the kernel.}
	\label{tab:occurrence}
    \vspace{-0.2in}
\end{table}

We also count the occurrences of individual  files involved in the  bug patches. 
Table~\ref{tab:occurrence} shows the top 10 most ``buggy'' files based on the occurrences  and the  average lines of code (LoC) changed per 100 LoC, 
which verifies that adding \texttt{dax}  and \texttt{nvdimm} supports are the two major sources of introducing PM bugs in the kernel. {In addition, the table also shows a fine-grained view (i.e., individual files) on where the bugs exist within a subsystem. Therefore, applying bug detection techniques (e.g., static code analysis) on these specific files and/or code paths may help identify most issues.}


%% file: sys_bugconseq.tex
\subsection{Bug Consequence}
\label{bug:bugconsequence}


To understand how severe the PM bugs are, 
we classify them based on the symptoms reported in the patches.
We find that there are 8 types of consequence, including  
 \textit{Missing Device}, \textit{Inaccessible Device}, \textit{Security},
\textit{Corruption}, \textit{Crash}, \textit{Hang}, \textit{Wrong Return Value}, and \textit{Resource Leak}.
We elaborate on the first four types 
as they are relatively more unique to our dataset, while the others are  similar to previous studies~\cite{Lu-FAST13-FS}:

First, \textit{Missing Device} implies the kernel is unable to detect PM devices, which is often the consequence of hardware dependent bugs. For example,  the \texttt{e820\_pmem} driver is responsible for registering resources that surface as \texttt{pmem} ranges.  However, the buggy `\texttt{e820\_pmem\_probe}'  method may fail to register the \texttt{pmem} ranges into the System-Physical Address (SPA) space, which makes the PM device not recognizable by the kernel.

Second, \textit{Inaccessible Device} means the PM device is detectable by the kernel but not accessible. For example, the `\texttt{start\_pad}' variable was introduced in  `\texttt{struct nd\_pfn\_sb}' of the \texttt{nvdimm} driver to record the padding size 
for aligning 
namespaces with the Linux memory hotplug section. 
But the buggy `\texttt{nd\_pfn\_validate}' method of the driver does not check for the variable, which leads to an alignment issue
and makes the namespace not recognizable by the kernel.

{Third, in terms of  \textit{Security}, we observe two interesting issues.
In one case,  write operations may be allowed
 on read-only \texttt{dax} mappings, 
which  exposes wrong access permissions to the end user. In another case, an NVDIMM's security attribute remains in `unlocked' state even when the admin issues an `overwrite' operation,
which could potentially allow malicious accesses to the PM device.
}

Fourth, \textit{Corruption} means the data or metadata stored on PM devices are corrupted, which could lead to permanent data loss if there is no additional backup available on other systems. One special type of corruption is the \textit{Crash-Consistency} issues, which means tricky corruptions (i.e., data or metadata inconsistencies) triggered by a crash event (e.g., power outage, kernel panic). Such issues have been investigated intensively in the literature due to its importance\cite{crashmonkey-hotstor17,Zheng-OSDI14-DB,Zheng-TOCS16-SSD}, and we have observed multiple crash-consistency issues in our dataset as well.
{For example,  the routine \texttt{dax\_mapping\_entry\_mkclean} may fail to clean and write protect DAX PMD entries; consequently,  memory-mapped writes to DAX PMD pages may not be flushed to PM even when a sync operation is invoked.  
As a result, a crash may leave the system in an inconsistent state.}



%% file: sys_bugfix.tex
\subsection{Bug Fix}
\label{sec:bugfix}
\begin{figure}[tb]
	\centering
	\includegraphics[width=3in]{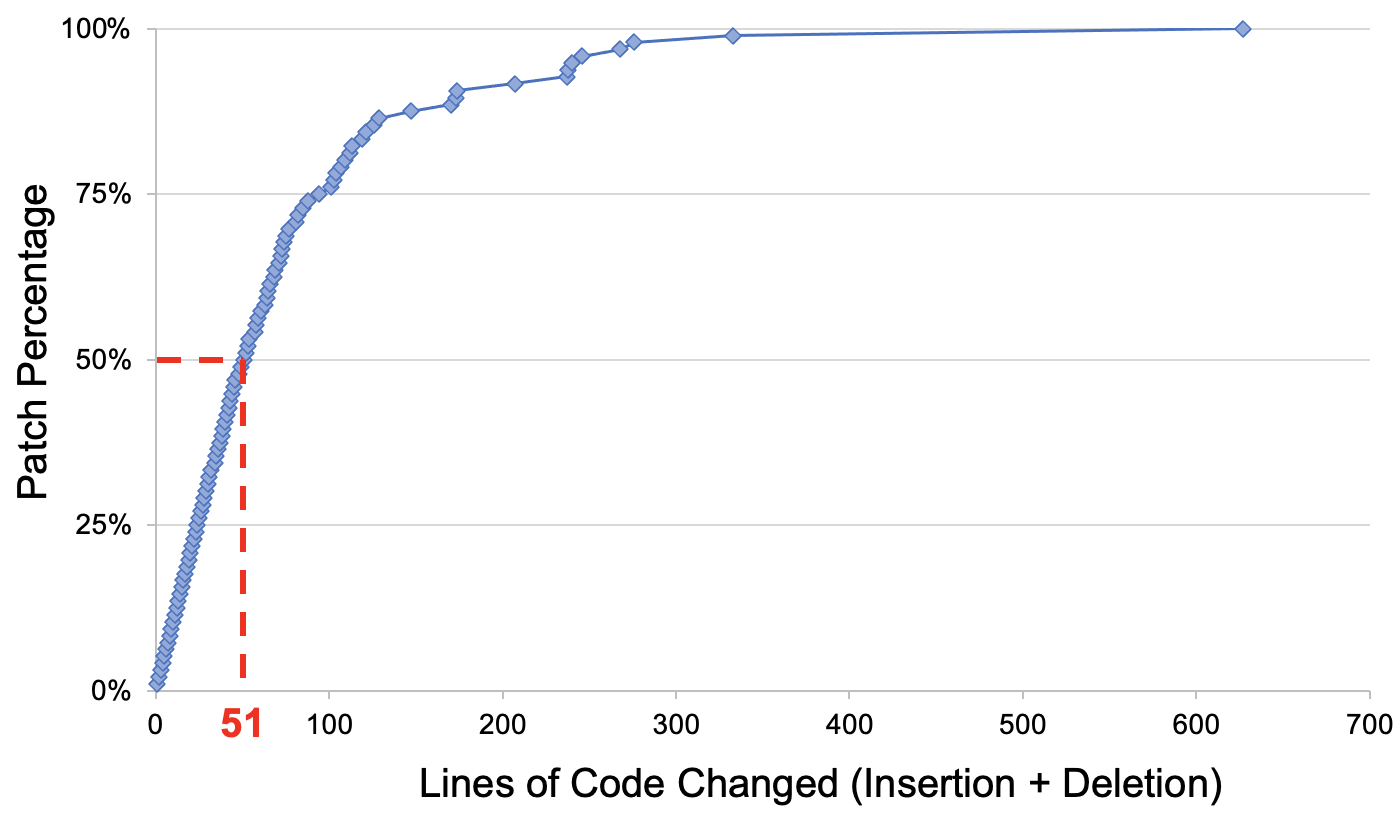}
	\caption{ {\bf {Size Distribution of PM Bug Patches}}}
	\label{fig:patchsize}
\end{figure}
\begin{table}[tb]
	\begin{center}
		\begin{tabular}{c|c|c|c|c|c|c}
			\textbf{File Count} & 1 &2 &3 & 4 & 5 & $>$ 5 \\
			\hline
			\textbf{Patch \%} & 66.0\% & 13.2\% & 8.3\% & 6.0\% & 2.9\% & 3.6\% \\
			\hline
			\hline
			\textbf{Dir. Count} & 1 &2 &3 & 4 & 5 & $>$ 5 \\
			\hline
			\textbf{Patch \%} & 79.2\% & 12.7\% & 5.7\% & 1.6\% & 0.5\% & 0.3\%  \\
		\end{tabular}
	\end{center}
	\caption{ {Scope of PM Bug Patches}. \textit{This table shows the \% of bug patch involving different counts of files or directories. }}
    \vspace{-0.3in}
	\label{tab:num_files_num_folfders}
\end{table}

\subsubsection{\textbf{How difficult it is to fix PM bugs}}
\label{sec:bugsize}

{Measuring the complexity of fixing PM bugs is challenging as it requires deep domain knowledge and may depend on the  developers' capability and other constraints (e.g., priority of tasks). Inspired by existing studies \cite{Lu-FAST13-FS,li2006-study,chou2001empirical}, we calculate three simple but quantitative metrics as follows, which might reflect the complexity to some extent:}

\vspace{0.05in}
\noindent
{\bf Bug Patch Size.} 
We define the patch size as the sum of lines of insertion and deletion  in the patch.
Figure \ref{fig:patchsize} shows the distribution of bug patch sizes.
We can see that most bug patches are relatively small.
For example, {50\% patches have less than 51 lines} of insertion and deletion code (LoC).
However, compared to traditional non-PM file system  bug patches where 50\% are less than 10 LoC~\cite{Lu-FAST13-FS}, 
the majority of  PM bug patches tend to be larger.



\vspace{0.05in}
\noindent
{\bf Bug Patch Scope.} We define the patch scope as the counts of files or directories involved in the patch.
For simplicity, we only count the top-level directories in the Linux source tree.
Table~\ref{tab:num_files_num_folfders} shows the patch scopes.
We can see that most patches only modified one file (66.0\%)  or files within one directory (79.2\%). 
On the other hand, 
3.6\% patches may involve more than 5 files.
Moreover, a non-negligible portion of  patches  involve  
more than one directories (20.8\%).
Since different directories represent different kernel subsystems, 
this implies that fixing these PM bugs are  non-trivial. 
For comparison, we randomly sample 100 GPU-related bug patches and measure their scope too.
We observe that only 5\% of the sampled GPU patches involve more than one directory, 
which is  much less than the 20.8\% cross-subsystem PM bug patches.  

{Although many PM bug patches in our dataset involves changes in multiple kernel subsystems, this does not  imply that only PM bug patches can involve multiple kernel subsystems. It is possible that there are complicated non-PM bugs which may also require  modifications across  multiple subsystems to fix. We leave further comparison of PM bugs and non-PM bugs in terms of the patch scope as future work.}



\vspace{0.05in}
\noindent
{\bf Time-to-Fix.} 
Most patches in our dataset do not contain the information \textit{when} the bug was first discovered. However, 
we find that 48 bug patches include links to the original bug reports, 
which enables us to measure the time-to-fix metric.
We find that PM bugs may take 6 to 48 days to fix with an average of 24 days,
which further implies the complexity. 
There are other sources which may provide more complete time-to-fix information (e.g., Bugzilla~\cite{KernelBugzillaWebsite}), which we leave as future work.
{Note that similar to measuring ``how long do bugs live'' in the classic study of Linux and OpenBSD kernel bugs~\cite{chou2001empirical},  human factors such as the developer's availability and capability may affect the time-to-fix metric; 
 therefore, there is unlikely a linear relationship between the time-to-fix metric and the bug complexity. In other words, how to measure the  complexity of bug precisely is still an open challenge}. On the other hand, 
we find that during the time to fix window, there were often in-depth discussions among developers which requires substantial domain knowledge and trial-and-error steps; therefore,
 we believe that the long time-to-fix value still reflects the need of better debugging support to certain extent, similar to the observations from  previous studies~\cite{li2006-study,chou2001empirical}.

\input{sys_fixstrategy.tex}

%% file: sys_fixstrategy.tex
\subsubsection{\textbf{Fix Strategy}}
\label{sec:fixstrategy}

\begin{figure}[t]
    \centering
    \begin{subfigure}[t]{0.5\textwidth}
        \centering
        \includegraphics[width=\textwidth]{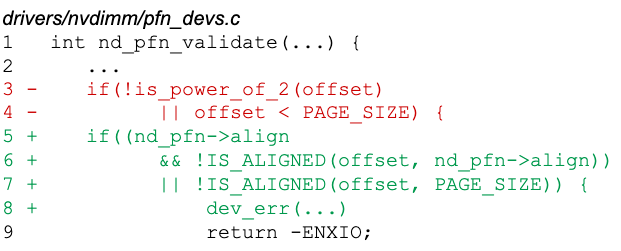}
        \vspace{-0.1in}
        \caption{{\bf An Alignment Bug Fix}}
    \label{fig:hw-sanity}
    \end{subfigure}
    ~
    \begin{subfigure}[t]{0.5\textwidth}
        \centering
        \includegraphics[width=\textwidth]{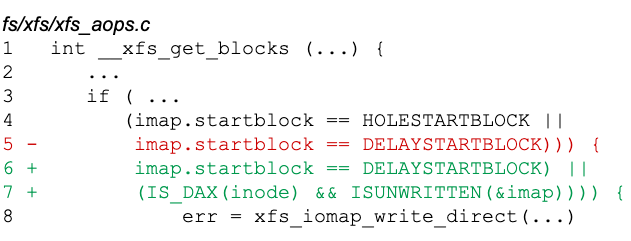}
        \vspace{-0.1in}
        \caption{{\bf A Data Race Bug Fix}}
    \label{fig:race-sanity}
    \end{subfigure}
    \caption{{Two examples of refining sanity checks to fix PM bugs: (a) Alignment bug fix; (b) Data race bug fix.}}
\end{figure}
We find that the strategies for fixing PM bugs often vary a lot depending on the specific bug types (Table~\ref{tab:bug-type}). 
On the other hand, we also  observe that different types of PM bugs  may be fixed by one common strategy: 
refining sanity checks. For example, {
Figure~\ref{fig:hw-sanity} shows an \textit{Alignment} bug fix.
In this example, a PM device was mistakenly disabled due to an ineffective sanity check (\texttt{is\_power\_of\_2}). 
The bug was fixed by replacing the original sanity check with an accurate one (\texttt{IS\_ALIGNED}). Similarly, Figure~\ref{fig:race-sanity} shows a data race bug that was triggered when there were concurrent write faults to same DAX pages. The error occurs because the DAX page fault was not aware of pre-allocated (unwritten data) and just-allocated blocks. Therefore, a data race may end up zeroing the pages, similar to the case described in Figure~\ref{fig:concurrency-bug} (\S\ref{sec:othertypes}). The bug was fixed in the filesystem layer by refining the sanity check to identify unwritten DAX pages. }
Similar fixes have been applied to other bugs triggered by check violations.

We also find that developers may \textit{temporarily} ``fix'' a PM bug by disabling a PM feature. 
For example, to avoid a race condition in handling transparent huge pages (THP) over  \texttt{dax},  
developers make the THP support over \texttt{dax} dependent on \texttt{CONFIG\_BROKEN}, 
 which means if \texttt{CONFIG\_BROKEN} is disabled (common case) then the feature is disabled too. 
 The developers even mention that a major re-work is required in the future, 
which implies the complexity of actually fixing the bug.

%% file: sys_reproduce.tex
\section{PM Bug Reproducibility}
\label{sec:reproduce}
\begin{table}[h]
    \centering
    \begin{tabular}{l|l|c}
        \textbf{Metric ID} & \hspace{27mm}\textbf{Description} & \textbf{Bug \#} \\
        \hline
        \textit{1Config} & Configuration parameters of PM device \& relevant software & 20\\
        \hline
        \textit{2HW} & Information of specific PM hardware type (e.g., NVDIMM-N) & 12\\
        \hline
        \textit{3Emu} & Information of PM emulation platform (e.g. QEMU) & 4\\
        \hline
        \textit{4Test} & Information of test suite and/or test case used (e.g. xfstests) & 18\\
        \hline
        \textit{5Step} & Information of necessary steps to reproduce the bug  & 18\\
       \hline
        \hline
        \multicolumn{2}{r|}{\textit{Total Unique}} & 39\\
    \end{tabular}
    \caption{{Five metrics for selecting PM bug candidates for reproducibility experiments.}}
    \label{tab:reproduce-params}
    \vspace{-0.3in}
\end{table}

To better understand the  conditions for manifesting the  issues and help derive effective remedy solutions,  we further perform a set of reproducibility experiments. We identify a subset of bug patches with relatively complete information, and 
attempt to reproduce them experimentally. 
We find that configuration parameters in different storage utilities (e.g., \texttt{mkfs} for creating a file system, \texttt{ndctl} for managing the \texttt{libnvdimm}  subsystem) are critically important for manifesting the issues, which suggests that it is necessary to take into account the configuration states when developing relevant testing or debugging tools. 



 As shown in Table \ref{tab:reproduce-params},
 we look for five pieces of  information from the bug patches when selecting candidates for  the reproducibility experiments, which includes configuration parameters (\textit{1Config}), hardware information (\textit{2HW}), emulation platform  (\textit{3Emu}), test suite information (\textit{4Test}), and reproducing steps information (\textit{5Step}). 
Based on the five metrics, we identify 39  bug candidates which contain relatively complete information for the experiments.

Table~\ref{tab:reproduce-bugs} summarizes the reproducibility results. At the time of this writing, we are able to reproduce 11 out of the 39 candidates based on the information derived from the patches (\cmark in the last column).
Most interestingly, we find that the configuration parameters of storage utilities (i.e., \texttt{mkfs}, \texttt{mount}, \texttt{ndctl}) and workloads are critically important for triggering the bugs, which are summarized in the four columns of "Critical Configurations".

For example, in bug \#15, when the file system is mounted with both DAX and read-only parameters, accessing the file system results in a ``Segmentation fault'', because the  DAX fault handler attempts to write to the journal in read-only mode.
As another example, the \texttt{ndctl} utility enables  configuring the \texttt{libnvdimm} subsystem of the Linux kernel. In bug  \#35, a resource leak was observed when an existing NVDIMM namespace was reconfigured to ``device-dax'' mode via  \texttt{ndctl}. 

Note that Table~\ref{tab:reproduce-bugs} only shows the necessary configurations for triggering the bug cases, which may not be sufficient.
In fact, there are many other factors which can make a case un-reproducible in our experiments. For example,
we are unable to reproduce cases that require test cases from the \texttt{ndctl} test suite~\cite{ndctl-github}. These test cases require ``nfit\_test.ko'' kernel module, which is not available in the generic Linux kernel source tree. 
Also, some cases  are dependent on specific architectures (e.g., PowerPC) which are incompatible with our experimental systems. In addition, many concurrency bugs  such as data races  cannot be easily reproduced due to 
the nondeterminism of thread interleavings.

Overall, we can see that most of the bug cases in Table \ref{tab:reproduce-bugs} require specific configurations to trigger, which suggests the importance of considering configurations for testing and debugging~\cite{SPEX,CTEST,CDEP,Taba-HotStorage22}. 
We summarize all the reproducible bug cases, including the necessary triggering conditions and scripts, in a public dataset to facilitate follow-up research~\cite{dsl-pmbug-repo}. 

%% file: sys_pmtools.tex
\begin{table}[ht]
    \centering
    \scriptsize
    \begin{tabular}{c|c|c|c|c|c|c}
        {\textbf{Bug}} &
        \multirow{2}{*}{\textbf{Bug Type}} & \multicolumn{4}{c|}{\textbf{Critical Configurations}} & \multirow{2}{*}{\textbf{R?}} \\
        \cline{3-6}
       {\textbf{ID}} &  & \textbf{mkfs} & \textbf{mount} & \textbf{ndctl} & \textbf{workload} & \\
        \hline
        1 & Semantic - State & - & - & - & \texttt{mmap(\tiny{MAP\_SHARED})} & \xmark \\
        \hline
        2 & Concurrency - Wrong lock & - & - & - & - & \xmark \\
        \hline
        3 & Error code - Error return & - & - & - & - & \xmark \\
        \hline
        4 & Semantic - Logic & - & \texttt{-o dax} & - & - & \cmark \\
        \hline
        5 & Semantic - Logic & - & \texttt{-o dax} & - & - & \cmark \\
        \hline
        6 & Semantic - Logic & - & - & - & - & \xmark \\
        \hline
        7 & Hardware - Alignment & - & - & \texttt{-a 4096} & - & \xmark \\
        \hline
        8 & Semantic - Logic & - & - & - & \texttt{fallocate} & \xmark \\
        & & & & & \tiny{\texttt{(FALLOC\_FL\_ZERO\_RANGE)}} & \\
        \hline
        9 & Hardware - Specification & - & - & - & - & \xmark \\
        \hline
        10 & Hardware - Cache & - & \texttt{-o dax} & - & - & \xmark \\
        \hline
        11 & Hardware - Alignment & - & - & \texttt{-m devdax} & - & \xmark \\
         & & & & \texttt{-a 4K} & & \\
        \hline
        12 & Semantic - Logic & - & - & \texttt{-m devdax} & - & \xmark \\
         & & & & \texttt{-a 4K} & & \\
        \hline
        13 & Concurrency - Race & - & \texttt{-o dax} & - & - & \xmark \\
        \hline
        14 & Semantic - Logic & - & \texttt{-o dax} & - & - & \cmark \\
        \hline
        15 & Semantic - Logic & - & \texttt{-o dax,ro} & - & - & \cmark \\
        \hline
        16 & Concurrency - Deadlock & - & \texttt{-o dax} & - & - & \xmark \\
        \hline
        17 & Semantic - Logic & \texttt{-O inline\_data} & \texttt{-o dax} & - & - & \cmark \\
        \hline
        18 & Semantic - Logic & - & \texttt{-o dax,} & - & - & \cmark \\
        & & & \texttt{nodelalloc} & & & \\
        \hline
        19 & Error code - Error check & - & \texttt{-o dax} & - & \texttt{open(\tiny{O\_TRUNC})} & \xmark \\
        \hline
        20 & Hardware - Alignment & - & - & \texttt{-m devdax} & - & \xmark \\
         & & & & \texttt{-a 1G} & & \\
         \hline
        21 & Semantic - Logic & \texttt{-t xfs} & \texttt{-o dax} & \texttt{-l 4K} & - & \xmark \\
        \hline
        22 & Semantic - Logic & - & - & - & \texttt{blockdev --setro} & \cmark \\
        \hline
        23 & Hardware - Specification & - & - & - & - & \xmark \\
        \hline
        24 & Semantic - Logic & - & - & - & - & \xmark \\
        \hline
        25 & Concurrency - Race &  \texttt{-t xfs} & \texttt{-o dax} & - & - & \xmark \\
        \hline
        26 & Concurrency - Race & - & - & - & - & \xmark \\
        \hline
        27 & Semantic - Logic & - & - & \texttt{-s < 16MB} & - & \xmark \\
        \hline
        28 & Semantic - Logic & - & - & \texttt{-e -m devdax} & - & \xmark \\
        \hline
        29 & Concurrency - Atomicity & - & - & \texttt{create-namespace} & - & \cmark \\
         & & & & \texttt{destroy-namespace} & & \\
        \hline
        30 & Concurrency - Deadlock & - & - & \texttt{create-namespace} & - & \cmark \\
         & & & & \texttt{destroy-namespace} & & \\
        \hline
        31 & Semantic - Logic & - & \texttt{-o dax} & - & \texttt{mmap(\tiny{MAP\_PRIVATE})} & \xmark \\
        \hline
        32 & Semantic - Logic & - & \texttt{-o dax} & - & \texttt{mmap(\tiny{MAP\_SYNC})} & \xmark \\
        \hline
        33 & Hardware - Compatibility & - & - & \texttt{-m devdax} & - & \xmark \\
         & & & & \texttt{-a 16M} & & \\
        \hline
        34 & Error code - Error return & - & - & \texttt{inject-error} & - & \xmark \\
        \hline
        35 & Memory - Resource leak & - & - & \texttt{-e -m devdax} & - & \cmark \\
        \hline
        36 & Semantic - Logic & - & - & \texttt{sanitize-dimm} & - & \xmark \\
         & & & & \texttt{--overwrite} & & \\
        \hline
        37 & Semantic - Logic & \texttt{-O inline\_data} & - & - & \texttt{chattr +x} & \cmark \\
        \hline
        38 & Semantic - Logic & - & - & \texttt{-e -m devdax} & - & \xmark \\
        \hline
        39 & Memory - Resource leak & - & - & - & - & \xmark \\
        \hline
    \end{tabular}
    \caption{Results of reproducing 39 PM bug cases. The last column \textit{R?} means if the bug case was reproducible (\cmark) or not (\xmark). The middle columns show the bug type as well as the critical configurations necessary for triggering the bug cases.}
    \label{tab:reproduce-bugs}
    \vspace{-0.3in}
\end{table}

\section{Implications on PM Bug Detection}
\label{sec:implications}

Our study has exposed a variety of PM-related issues, 
which may help develop effective PM bug detectors and build robust PM systems. 
{
For example, since 20.8\%  PM bug patches involve multiple
kernel subsystems, simply focusing on one subsystem is unlikely
enough; instead, a full-stack approach is much needed, and identifying the potential dependencies among components would be critical. On the other hand, since many bugs in different subsystems may follow similar patterns, capturing one bug pattern may
benefit multiple subsystems (see \S\ref{sec:importance} for more discussions).
}

{As one step to address the PM-related issues identified in the study,  we analyze a few state-of-the-art bug detection tools in this section. We  discuss the limitations as well as the opportunities for both PM bug detectors and Non-PM bug detectors 
(\S\ref{sec:pmbugtools} and \S\ref{sec:nonpmbugtools}, respectively).  Moreover (\S\ref{sec:drchecker})}, we present our efforts and results on extending one state-of-the-art static bug detector (i.e., Dr. Checker~\cite{drchecker}) for analyzing PM drivers,   which account for the majority of bug cases in our dataset (Figure~\ref{fig:folders_num_files2}).

\subsection{PM Bug Detectors}
\label{sec:pmbugtools}

{Multiple PM-specific bug detection tools have been proposed recently, including PMTest~\cite{liu2019pmtest}, XFDetector~\cite{liu2020cross}, and AGAMOTTO~\cite{AGAMOTTO-OSDI20}. These tools mostly focus on user-level PM programs. We have performed bug detection experiments using these tools, and we are able to verify their effectiveness by reproducing most of the bug detection results reported in the papers.}
Unfortunately, 
we find that they are fundamentally limited for capturing the PM bugs in our dataset. 
For example, XFDetector~\cite{liu2020cross} relies on Intel Pin~\cite{pin} which can only instrument user-level programs.  
PMTest~\cite{liu2019pmtest} can be applied to kernel modules, but it requires manual annotations which is impractical for major kernel subsystems.  
AGAMOTTO~\cite{AGAMOTTO-OSDI20} relies on KLEE~\cite{KLEE} to 
symbolically explore user-level PM programs.  
While it is possible to integrate KLEE with virtual machines to enable full-stack symbolic execution (as in S2E~\cite{S2E}), novel PM-specific path reduction algorithms are likely needed to avoid the state explosion problem~\cite{state-merge-pldi12}. 
One recent work Jaaru~\cite{gorjiara2021jaaru} leverages commit stores, a
common coding pattern in user-level PM programs, to reduce the number of execution paths  that need to be explored for model checking. 
Nevertheless,  such elegant pattern has not been observed in our dataset due to the complexity of kernel-level PM optimizations (\S\ref{sec:pmbugs}). Therefore, additional innovations on path reduction are likely needed to apply  model checking to detect  diverse PM-related bugs in the kernel effectively.

\begin{table}[t]
    \centering
    \small
    \begin{tabular}{l|p{0.65\linewidth}|c | c}
        \textbf{Driver Module} & \hspace{3cm}\textbf{Description} & \textbf{Ck} & \textbf{Ck+} \\
        \hline
        nfit.ko & Probe NVDIMM devices and register a libnvdimm device tree. Enables libnvdimm driver to pass device specific messages (DSM) for platform/DIMM configuration. & \xmark{}  & \cmark{}\\
        \hline
        nd\_pmem.ko & Drives a system-physical-address (SPA) range where memory store operations are persisted. Enables support for Direct-Access (DAX) feature. & \xmark{}  & \cmark{}\\
        \hline
        nd\_blk.ko & Enables I/O access to DIMM over a set of programmable memory mapped apertures. 
        A set of apertures can access just one DIMM device. & \xmark{}  & \cmark{}\\
        \hline
        nd\_btt.ko & Enables an indirection table that provides power-fail-atomicity of at least one sector (512B). Can be used in front of PMEM or BLK block device drivers. & \xmark{}  & \cmark{}\\
        \hline
        libnvdimm.ko & Provides generic support for NVDIMM devices, such as discover NVDIMM resources, register and advertise PM namespaces (e.g. /dev/pmemX, /dev/daxX.X, etc.) & \cmark{}  & \cmark{} \\
        \hline
        device\_dax.ko & Support raw access to PM over an mmap capable character device. & \xmark{}  & \cmark{}\\
        \hline
        dax.ko & Provides generic support for direct access to PM. & \xmark{}  & \cmark{}\\
        \hline
        
    \end{tabular}
    \caption{List of major PM driver modules studied in this work. The last two columns   show whether the module can be  supported  (\cmark{}) or not (\xmark{}) by  the vanilla Dr.Checker (\textbf{Ck}) and our extended version Dr.Checker+ (\textbf{Ck+}). The vanilla Dr.Checker can only  support the  \texttt{ioctl} entry type used in \texttt{libnvdimm.ko}.
    }
    \label{tab:nvdimm-drivers}
    \vspace{-0.3in}
\end{table}

\subsection{Non-PM Bug Detectors} 
\label{sec:nonpmbugtools}

Great efforts have been made to detect non-PM bugs in the kernel~\cite{crashmonkey-hotstor17,janus-ieee-sp19,hydra-sosp19,razzer-ieee-sp19,krace-ieee-sp20,KASAN,UBSan,Kmemcheck}. 
For example, CrashMonkey~\cite{crashmonkey-hotstor17} logs the \texttt{bio} requests and emulates crashed disk states
to test the crash consistency of traditional file systems. As discussed in \S\ref{bug:bugconsequence}, such tricky crash
consistency issues  exist in PM subsystems too. Nevertheless, extending CrashMonkey to detect PM
bugs may require substantial modifications including tracking PM
accesses and PM-critical instructions (e.g., mfence), designing PM-specific workloads, among others.


 Similarly, fuzzing-based tools have proven to be effective for kernel bug detection~\cite{syzkaller,janus-ieee-sp19,hydra-sosp19,razzer-ieee-sp19,krace-ieee-sp20}. 
For example, Syzkaller \cite{syzkaller} is a kernel fuzzer that executes kernel code paths by randomizing inputs for various system calls and has been the foundation for building  other fuzzers;
Razzer~\cite{razzer-ieee-sp19} combines fuzzing with static analysis 
 and detects data races  in multiple kernel subsystems (e.g., `\texttt{driver}', `\texttt{fs}', `\texttt{mm}'),
 which could potentially be extended to cover a large portion of concurrency PM bugs in our dataset. Since Syzkaller, Razzer and similar fuzzers heavily rely on virtualized  (e.g., QEMU~\cite{qemu}) or simplified (e.g., LKL~\cite{LKL}) environments to achieve high efficiency for kenel fuzzing, 
 one common challenge and opportunity for extending them is to 
 emulate PM devices and  interfaces precisely to ensure the fidelity.

Also, Linux kernel developers have incorporated tools such as Kernel Address Sanitizer (KASAN) ~\cite{KASAN}, Undefined Behavior Sanitizer (UBSan)~\cite{UBSan} and memory leak detectors (Kmemcheck)~\cite{Kmemcheck} within the kernel code to detect various memory bugs (e.g., null pointers, use-after-free, resource leak). These sanitizers  instrument the kernel code during compilation and examine bug patterns at runtime.
Similar to other dynamic tools, these tools can only detect issues on the executed code paths.  In other words, their effectiveness heavily depends on the quality of the inputs. 
As discussed in \S\ref{sec:reproduce}, many PM issues in our dataset require specific configuration parameters from utilities (e.g., \texttt{mkfs}, \texttt{ndctl}) and  workloads (e.g., \texttt{mmap(MAP\_SHARED)}, \texttt{open(O\_TRUNC)})  to trigger, so we believe it is important to consider such PM-critical configurations when leveraging existing kernel sanitizers for detecting the issues exposed in our study.

As discussed above, while various bug detectors have been proposed and used in practice, addressing  PM-related issues in our dataset will likely require PM-oriented innovations including precise PM emulation, PM-specific configuration and workload support, etc., which we leave as future work. 

On the other hand, to better understand the feasibility of extending existing tools for PM bug detection, we present our efforts and results on extending one existing bug detector called Dr.Checker~\cite{drchecker}  in this section. We select Dr.Checker for {three} main reasons: First, {it has proven 
effective } for analyzing kernel-level drivers~\cite{drchecker}, and as shown in Figure~\ref{fig:folders_num_files2} (\S\ref{sec:where}), drivers account for the majority of bugs in our dataset; Second, Dr.Checker is based on static analysis without dynamic execution, which makes it less sensitive to the limitations discussed in the previous sections (e.g., device emulation, input generation); 
{Third, Dr. Checker employs multiple bug detection algorithms that can detect multiple bug patterns identified in our study (\S\ref{sec:bugtype}).}
We name our extension as Dr.Checker+ and release it on Git to facilitate follow-up research on PM bug detection~\cite{dsl-pmbug-repo}.

\begin{figure}[tb]
	\centering
	\includegraphics[width=3.5in]{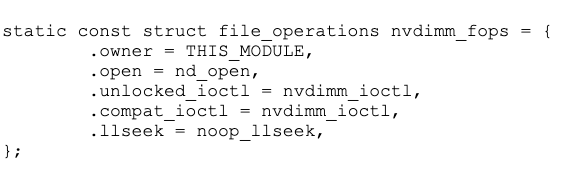}
	\vspace{-0.2in}
 \caption{{Entry points for   the \texttt{libnvdimm} driver module}}
	\label{fig:entrypoint}
	 \vspace{-0.2in}
\end{figure}

\smallskip
\noindent
\textbf{About Dr.Checker \& its limitations.} Dr.Checker~\cite{drchecker} mainly uses two static analysis techniques (i.e., points-to and taint analysis) to detect memory-related bugs (e.g., buffer overflow) in generic Linux drivers.  It performs flow-sensitive and context-sensitive code traversal to achieve high precision for driver code analysis.
One special requirement for applying  Dr.Checker is to identify correct \textit{entry points} (i.e.,  functions invoking the driver code) as well as the  argument types of entry points. 
{
For example, Figure~\ref{fig:entrypoint} shows a VFS interface (``struct file\_operations'') with function pointers that allow userspace programs to invoke  operations on \texttt{libnvdimm} driver module. The functions included in the structure (e.g., ``\texttt{nvdimm\_ioctl}'') are entry points  
that enables manipulating the underlying NVDIMM devices through the driver code. }
The entry points and their argument types  collectively determine the \textit{entry types}, which in turn determines the  \textit{taint sources} for initiating relevant analysis. {For example, the vanilla Dr. Checker describes an IOCTL \textit{entry type}  where the first argument of the entry point function is marked as \textit{PointerArgs} (i.e., the argument points to a kernel location, which contains the tainted data) and the second argument is marked as \textit{TaintedArgs} (i.e., the argument contains tainted data and is referenced directly). This entry type is applicable to entry point functions that have similar signature. 
In the case of the  \texttt{libnvdimm} kernel module, 
the IOCTL entry type is applicable to two entry point functions (i.e., \texttt{nd\_ioctl}, \texttt{nvdimm\_ioctl}).
}
In addition, Dr.Checker includes a number of \textit{detectors} to check specific bug patterns based on the taint analysis
(e.g., \textit{IntegerOverflowDetector}, \textit{GlobalVariableRaceDetector}).   

While Dr.Checker has been applied to analyze a number of  Linux device drivers~\cite{drchecker}, it does not  support major PM drivers directly. 
{
Table~\ref{tab:nvdimm-drivers} summarizes the seven PM driver modules involved in our study, including \texttt{nfit.ko}, \texttt{nd\_pmem.ko}, \texttt{nd\_blk.ko}, \texttt{nd\_btt.ko}, \texttt{libnvdimm.ko}, \texttt{device\_dax.ko}, and \texttt{dax.ko}. These driver modules  provide various supports to make PM devices usable on Linux-based systems. For example, \texttt{dax.ko} provides generic support for direct access to PM, which is critical for implementing the {dax} feature for Linux file systems including Ext4 and XFS. As discussed in \S\ref{sec:pmbugs} and \S\ref{sec:reproduce}, these PM drivers tend to contain the most PM bugs in our dataset, and many PM bugs require PM driver features to trigger (e.g., \texttt{-o dax}). Based on our study of the bug patterns and relevant source code, we}  find that these PM drivers have much more diverse entry types compared to the embedded drivers analyzed by the vanilla Dr.Checker~\cite{drchecker}. As shown in the second to the last column of Table~\ref{tab:nvdimm-drivers}, the vanilla Dr.Checker~\cite{drchecker} only supports one entry type (i.e., \texttt{ioctl(file))} used in the \texttt{libnvdimm} module,  leaving the majority  of PM driver code unattended. 

\smallskip
\noindent
\textbf{Extending Dr.Checker to Dr.Checker+.} To make Dr.Checker works for the major PM drivers, 
we manually examine the source code of PM drivers and identify  critical entry points. As summarized in Table~\ref{tab:entry-types}, 
we have added five new  entry types (i.e., DEV\_OP, DAX\_OP, BLK\_OP, GETGEO, MOUNT) besides the original IOCTL entry type. The new entry types include the major entry point functions used in the PM driver modules. Moreover, we identify the critical input arguments and map them to the  appropriate taint types defined by Dr.Checker (i.e., \textit{TaintedArgs} for arguments directly passed by the userspace, and \textit{PointerArgs} and \textit{TaintedArgsData} for arguments pointing to a kernel memory location which contains tainted data).
In addition, to make Dr.Checker work for newer versions of Linux kernel, we have ported the implementation to the latest version of the LLVM/Clang compiler infrastructure~\cite{LLVM}. 
{Overall, as summarized in the last column of Table~\ref{tab:nvdimm-drivers}, the enhanced Dr. Checker+ is able to support bug detection in the seven major PM driver modules based on our comprehensive study of the PM driver bugs and the relevant source code.}

\subsection{Extending Dr. Checker for Analysing PM Kernel Modules}
\label{sec:drchecker}
\begin{table}[t]
    \centering
    \scriptsize
    \begin{tabular}{c|c|l|c|c}
        \textbf{Entry} & \textbf{Kernel} & \multirow{2}{*}{\hspace{7mm}\textbf{Entry Point Function(s)}} & \multirow{2}{*}{\textbf{Tainted Input Argument (s)}} & \multirow{2}{*}{\textbf{Taint Type}} \\
        \textbf{Type} & \textbf{Module} &  & & \\
        \hline
        \multirow{7}{*}{DEV\_OP*} & \multirow{2}{*}{nfit.ko} &\texttt{acpi\_nfit\_add}, \texttt{acpi\_nfit\_probe} & \multirow{2}{*}{\texttt{struct acpi\_device *}\textit{adev}} & \multirow{2}{*}{PointerArg} \\
        & & \texttt{acpi\_nfit\_remove} & & \\
        \cline{2-5}
        & nd\_blk.ko & \texttt{nd\_blk\_probe}, \texttt{nd\_blk\_remove}
        & \multirow{2}{*}{\texttt{struct device *}\textit{dev}} & \multirow{2}{*}{PointerArg} \\
        \cline{2-3}
        & libnvdimm.ko & \texttt{nvdimm\_release}, \texttt{nvdimm\_probe}& &\\
        \cline{2-5}
        &\multirow{2}{*}{device\_dax.ko} & \texttt{dev\_dax\_split} & \texttt{struct vm\_area *}\textit{vma} & PointerArg\\
        \cline{3-5}
        & & \texttt{dev\_dax\_fault}, \texttt{dev\_dax\_pagesize} & \texttt{struct vm\_fault *}\textit{vmf} & PointerArg\\
        \cline{2-5}
        & dax.ko & \texttt{dax\_destroy\_inode}, \texttt{dax\_free\_inode} & \texttt{struct inode *}\textit{inode} & PointerArg\\
        \hline
        \multirow{3}{*}{DAX\_OP*} & \multirow{3}{*}{nd\_pmem.ko} & \multirow{3}{*}{\texttt{pmem\_dax\_direct\_access}} & \texttt{struct dax\_device *}\textit{dax\_dev} & PointerArg\\
        \cline{4-5}
        & & & \texttt{pgoff\_t }\textit{pgoff} & TaintedArgs\\
        \cline{4-5}
        & & & \texttt{long }\textit{nr\_pages} & TaintedArgs\\
        \hline
        \multirow{4}{*}{BLK\_OP*} & \multirow{2}{*}{nd\_pmem.ko} & \multirow{2}{*}{\texttt{pmem\_rw\_page}} & \texttt{struct block\_device *}\textit{bdev} & PointerArg\\
        \cline{4-5}
        & & & \texttt{sector\_t }\textit{sector} & TaintedArgs\\
        \cline{2-5}
        & \multirow{2}{*}{nd\_btt.ko} & \multirow{2}{*}{\texttt{btt\_rw\_page}} & \texttt{struct page *}\textit{page} & TaintedArgsData\\
        \cline{4-5}
        & & & \texttt{unsigned int }\textit{op} & TaintedArgs\\
        \hline
        GETGEO* & nd\_btt.ko & \texttt{btt\_getgeo} & \texttt{struct block\_device *}\textit{bdev} & TaintedArgs\\
        \hline
                MOUNT* & dax.ko & \texttt{dax\_mount} & \texttt{struct file\_system\_type *}\textit{fs\_type} & TaintedArgsData\\
        \hline
        \multirow{2}{*}{IOCTL} & \multirow{2}{*}{libnvdimm.ko} & \multirow{2}{*}{\texttt{nd\_ioctl, nvdimm\_ioctl}} & \texttt{struct file *}\textit{file} & PointerArgs\\
        \cline{4-5}
        & & & \texttt{unsigned long }\textit{arg} & TaintedArgs\\
        \hline
    \end{tabular}
    \caption{Entry types identified to support Dr.Checker for analyzing PM drivers. The first five types (*) are newly added entry types. The table also shows the specific functions, arguments, and taint types applied for individual arguments.}
    \label{tab:entry-types}
    \vspace{-0.3in}
\end{table}

\smallskip
\noindent
\textbf{Experimental Results of Dr.Checker+.}
We have applied the extended Dr.Checker+ to analyze seven  major PM kernel modules in Linux kernel v5.2.
{In this set of experiments, we applied four  {detectors}: (1)\textit{IntegerOverflowDetector} checks for tainted data used in operations (e.g., \texttt{add}, \texttt{sub} or \texttt{mul}) that may cause an integer overflow or underflow; (2) \textit{TaintedPointerDereferenceChecker} detects pointers that are tainted and directly dereferenced; (3) \textit{GlobalVariableRaceDetector} checks  for global variables that are accessed without a mutex; and (4) \textit{TaintedSizeDetector} checks for tainted data that is  used  as  a  size  argument  in  any  of  the \texttt{copy\_to\_} or \texttt{copy\_from\_} functions which may result in information leak or buffer overflows.
}
Table~\ref{tab:dr_checker_results} summarizes the experimental results.
 Overall, Dr. Checker+ can process all the target kernel modules successfully, and we have observed  warnings (i.e., potential issues) reported by its four detectors in five out of the seven kernel modules (i.e., \texttt{nfit.ko},  \texttt{nd\_pmem.ko},  \texttt{nd\_btt.ko},  \texttt{libnvdimm.ko}, and  \texttt{dax.ko}).  {For example, the \textit{TaintedPointerDereferenceChecker} was able to identify a potential null pointer dereference in the   \texttt{nd\_btt} driver module, where the pointer variable \texttt{bdev} in the \texttt{btt\_getgeo} entry point of  \texttt{nd\_btt} was accessed without a check for its validity. If the routines invoking the entry point function do not check for the null value before using it, then this code may lead to a kernel crash.}


Note that the warnings reported by Dr.Checker+ do not necessarily imply PM bugs due to the conservative static analysis used in Dr.Checker. For example, the \textit{GlobalVariableRaceDetector} in Dr. Checker would falsely report a warning for any access to a global variable outside of a critical section. By looking into the warnings reported in our experiments, we observed a similar false alarm: Dr. Checker+  may report a warning when the driver code invokes the macro "WARN\_ON" to print to the kernel error log, which is benign.  

Also, since Dr.Checker's detectors are stateless, they may report a warning for every occurrence of the same issue. For example, the page offset in \texttt{pmem\_dax\_direct\_access} is used to calculate the physical address of a page, which involves bit manipulation operations and the resulting value may overflow the range of possible integer values. The \textit{IntegerOverflowDetector} may report a warning whenever the method is invoked.

\begin{table}[t]
    \centering
    \footnotesize
    \begin{tabular}{c|c|c|c|c|c|c}
        \multirow{2}{*}{\bf Detectors of Dr.Checker+} & \multicolumn{5}{c|}{\bf{PM Kernel Modules}} & \multirow{2}{*}{\bf Total} \\
        \cline{2-6}
         & \footnotesize{\bf{nfit.ko}} & \footnotesize{\bf{nd\_pmem.ko}} & \footnotesize{\bf{nd\_btt.ko}} & \footnotesize{\bf{libnvdimm.ko}} & \footnotesize{\bf{dax.ko}} & \\
         \hline
         IntegerOverflowDetector & 0 & 4 & 2 & 0 & 0 & 6 \\
         \hline
         TaintedPointerDereferenceChecker & 0 & 0 & 4 & 0 & 0 & 4 \\
         \hline
         GlobalVariableRaceDetector & 1 & 0 & 0 & 12 & 2 & 15 \\
         \hline
         TaintedSizeDetector & 0 & 0 & 0 & 4 & 0 & 4\\
         \hline
         \hline
         \bf{Total} & 1 & 4 & 6 & 16 & 2 & 29\\
         \hline
    \end{tabular}
   \caption{Summary of warnings reported by Dr. Checker+ on analyzing five PM kernel modules. 
   }
    \label{tab:dr_checker_results}
    \vspace{-0.3in}
\end{table}

Overall, our experience is that extending Dr.Checker for analyzing PM-related issues in the Linux kernel is non-trivial and time-consuming due to the complexity of  the PM subsystem. On the other hand, the actual code modification is minor: we only need to modify about 100 lines of code (LOC) in Dr.Checker to cover the major PM driver modules. 
While the effectiveness is still fundamentally limited by the capability of the core techniques used in the existing tools (e.g., Dr.Checker's static analysis may report false alarms), we believe that extending existing tools to make them work for the PM subsystem in the kernel can be an important   first step towards  addressing the PM-related challenges exposed in our study. {We leave the investigation of improving Dr.Checker further (e.g., improving the static analysis algorithms, adding diagnosis support for understanding the root causes of warnings, or fixing the warnings detected) and building new detection tools as future work}.

\color{black}

%% file: related.tex
\section{Related Work}
\label{sec:related}



\noindent
{\bf Studies of Software Bugs.}
Many researchers have performed empirical studies on bugs in open source software~\cite{lu2008learning,Lu-FAST13-FS,chou2001empirical,lazar2014does,chen2011linux,gunawi2014bug,zhang2022reproducibility,Taba-HotStorage22,Taba-FAST23,GRace-ppopp11,asplos11-2ndstrike,igsc17-edelta}. 
For example, Lu \etal{}~\cite{lu2008learning} studied 105 concurrency bugs from 4 applications
 and found that atomicity-violation and order-violation are two common bug patterns; 
 Lu \etal{}~\cite{Lu-FAST13-FS} studied 5,079 file system patches (including 1,800 bugs fixed between Dec. 2003 and May 2011) and identified the  trends of 6 file systems; Mahmud \etal{}~\cite{Taba-HotStorage22,Taba-FAST23} studied bug patterns in file systems and utilities and extracted configuration dependencies that may affect the manifestation of bugs. 
Our study is complementary to the existing ones as we focus on bugs related to the latest PM technology, 
 which may involve issues beyond existing foci (e.g., user-level concurrency bugs~\cite{lu2008learning,asplos11-2ndstrike,gmrace-tpds14,GRace-ppopp11}, non-PM file systems~\cite{Lu-FAST13-FS},   cryptographic modules~\cite{lazar2014does}, configurations~\cite{Taba-HotStorage22,Taba-FAST23}).  

\vspace{0.05in}
\noindent
{\bf Studies of Production System Failures.}
Researchers have also studied  
various failure incidents in production systems~\cite{gunawi2016does,xu2019lessons,liu2019bugs,guo2013failure,gunawi2018fail,xu2018understanding,zhang2021benchmarking,TOS22-Runzhou,pdsw20-fingerprinting,hpec18-yehia}, many of which are caused by software bugs.
For example, Gunawi \etal{}~\cite{gunawi2016does} studied 
597 cloud service outages and derived multiple lessons including the outage impacts, causes, etc.; they found that many root causes were not described clearly. 
Similarly, Liu \etal{}~\cite{liu2019bugs} studied hundreds of high-severity 
incidents  in Microsoft Azure. 
Due to the nature of the data source, these studies 
typically focus on high-level insights 
(e.g., caused by hardware, software, or human factors) 
instead of source-code level bug patterns described in this study.
Since PM-based servers are emerging for production systems~\cite{delldcpmm},  and many production systems are based on Linux kernel,
our study may help understand PM-related incidents in the real world.

\vspace{0.05in}
\noindent
{\bf Tools for Testing \& Debugging Storage Systems.} Many tools have been created to test storage systems~\cite{Yang-OSDI06-EXPLODE,Changwoo-SOSP15-CrosscheckingFS,vjay-osdi18-blackboxtesting,Zheng-OSDI14-DB,Janus,zheng2013understanding,Om-FAST18-RFSCK,Taba-FAST23,TOS22-Runzhou,ics18-pfault,Cao-PDSW16} or help debug system failures~\cite{zhang2021benchmarking,FailureSketchSOSP15,OmniTableHotOS19,sentilog,hpdc22provio,ipdps23drill,igsc17-checkpoint,ipdps23faultyrank}. For example, EXPLODE~\cite{Yang-OSDI06-EXPLODE}, B$^3$~\cite{vjay-osdi18-blackboxtesting}, and Zheng et.al. ~\cite{Zheng-OSDI14-DB}  
apply fault injections to emulate crash images
to detect crash-consistency bugs in file systems, which has also been observed in our dataset.
PFault ~\cite{ics18-pfault,TOS22-Runzhou} applies fault injection to test Lustre and BeeGFS parallel file systems building on top of the Linux kernel.
Gist~\cite{FailureSketchSOSP15} applies program slicing to help pinpoint the root cause of failures in storage software including SQLite, Memcached, etc. Duo et.al.~\cite{zhang2021benchmarking} extended PANDA~\cite{panda} to track device commands to help diagnosis.
In general, these tools rely on detailed understanding of the underlying bug patterns  to be effective. We believe our study on PM-related issues can contribute to building more effective bug detectors or debugging tools for PM-based storage systems, which we leave as future work.

%% file: sys_discussion.tex
\section{Discussions}
\label{sec:discussion}
\subsection{Importance of Empirical Studies on Real-World Bug Patterns}
\label{sec:importance}
{As briefly mentioned in previous sections, this work was inspired by  many existing research efforts on studying the characteristics of bugs in real-world software systems, including  bug patterns in multi-threaded applications~\cite{lu2008learning},  
Linux and/or OpenBSD kernels~\cite{chou2001empirical,chen2011linux},   (non-PM) file systems~\cite{Lu-FAST13-FS} and utilities~\cite{Taba-HotStorage22},   cryptographic software~\cite{lazar2014does},  cloud systems~\cite{gunawi2014bug}, and so on.
Due to the complexity of real-world systems, it is practically impossible to build effective or efficient tools to address the issues induced by various bugs  without thorough understanding of the bug characteristics.
Therefore, such empirical studies of real-world bug patterns, which typically requires substantial manual efforts and domain knowledge,  often serve as the foundation for understanding the issues and deriving practical solutions later. 
For example, the seminal work of S. Lu et. al.~\cite{lu2008learning} studied the concurrency bug patterns from four applications; while the study did not  build any tools to address the bugs directly, it has inspired a series of follow-up research on concurrency bug detection and improving multi-threaded software in general~\cite{lu2011detecting,wen2022controlled,deng2013efficient,gao20112ndstrike}. Similarly, the study of Linux file system patches by L. Lu et. al. ~\cite{Lu-FAST13-FS} has exposed general bug patterns in the context of file systems and has inspired various follow-up innovations on improving file systems reliability~\cite{Changwoo-SOSP15-CrosscheckingFS,Om-FAST18-RFSCK}. One common lesson learned from these classic works is that ``history repeats itself'', and ``learning from mistakes'' is important for a better future~\cite{lu2008learning,Lu-FAST13-FS}.}

{Our study follows a similar methodology (\S\ref{sec:methodology}) but focus on a dataset that has never been covered by existing works (to the best of our knowledge). As discussed in \S\ref{sec:pmbugs},  we have identified a wide range of PM-related bug patterns, some of which  are consistent to previous findings but others are not. For example,  our study shows that cache-related issues 
 only contribute to a small subset (1.7\%) of the bugs in our dataset, and crash-consistency issues may  be caused by semantic and concurrency bugs besides misusing low-level instructions.  This observation is  
in contrast to   recent research focus on PM bug detection in the  community which mostly only consider crash-consistency issues caused by the misuse of cacheline flush instructions~\cite{AGAMOTTO-OSDI20,liu2020cross,liu2019pmtest,gorjiara2021jaaru}. 
In other words,  additional research efforts are likely needed to address the PM-related issues in general.} 

{As exemplified by previous studies, we envision  multiple directions for follow-up research based on our empirical study of PM bug patches. First,  the detailed characterisation of PM bug patterns (\S\ref{sec:pmbugs}) may help derive concrete rules for building new PM bug detection tools, similar to our extension to Dr. Checker (\S\ref{sec:drchecker}). Second,  the critical conditions identified in our reproducibility experiments (\S\ref{sec:reproduce}) may guide the design of future tools which need to consider how to trigger the bugs for bug detection or failure diagnosis. Third, the curated dataset of various types of PM bugs may serve as the test cases for evaluating the effectiveness of newly built tools, similar to the concept of BugBench~\cite{lu2005bugbench,zhang2021benchmarking}. We hope that by publishing our study results and releasing the dataset, the work can facilitate building the next  generation of
more robust PM-based  storage systems.}


\subsection{Importance beyond Intel Optane DCPMM}

{As mentioned in {\S\ref{sec:intro}} and {\S\ref{sec:unique}}, PM technologies are not limited to Intel  Optane{\textsuperscript{\texttrademark}} DCPMM~{\cite{IntelDCPM}}, and the  PM ecosystem in Linux is not solely designed for Intel.   
Various other PM technologies 
(e.g., PCM~{\cite{pcm}}, STT-RAM~{\cite{STTRAM}}, CXL-based PM~{\cite{cxl-pmem,bhardwaj2022cache}}) will likely require similar  support from the Linux kernel (e.g., the {\texttt{dax}} feature). Therefore, our study of PM-related issues in the Linux kernel is not limited to Intel Optane technology either.}

{While Intel is winding down its Optane DCPMM business and will only provide limited support in the next  3 to 5 years~\cite{intel-pmemio-update}, we expect to see other PM devices in the near future which will be largely compatible with the Linux PM ecosystem.
Particularly, 
Compute Express Link (CXL) \cite{cxl-consortium} is a new open standard cache-coherent interconnect for processors, memory expansion and accelerators. 
CXL 2.0 specification describes that PM devices can be realized using CXL.io and CXL.mem protocols \cite{cxl-pmem}. 
Hence, it is expected that future PM devices would reside on CXL slots instead of memory DIMM slots. 
Although this change may impact how the host system recognizes PM devices, it is expected to have little impact on the PM software stack. 
For example, CXL-PM driver code in the Linux kernel (\texttt{drivers/cxl/pmem.c}) relies on existing NVDIMM driver code for functions such as \textit{register}, \textit{un-register} and \textit{probe} devices \cite{cxl-pmem-linux}.
CXL-based PM devices would still appear as special memory devices (similar to existing PM devices), and existing software interfaces (e.g., DAX) are expected to work without any modifications. 
Therefore, our study on the current Linux PM subsystem, which is the foundation for supporting  the next generation of  CXL-based PM devices, is still relevant. On the other hand, we expect that CXL-based devices will likely introduce new CXL-specific drivers and specifications,  which will add to the complexity of Linux PM subsystem and may introduce new bug patterns. Therefore, additional studies will likely be needed in the future.
}

%% file: conclude.tex
 \section{Conclusions and Future Work}
 \label{sec:conclude}
 
This paper presented a comprehensive study on PM-related patches and bugs in the Linux kernel. Based on 1,553 PM-related kernel patches, 
we derived multiple insights in terms of patch categories, bug patterns, consequences, fix strategies, etc. 
We also performed reproducibility experiments to identify the critical configuration conditions for manifesting the bug cases. Moreover, we conducted tool extension experiments to explore the remedy solutions to address the issues exposed in our study.
{In the future, we plan to investigate new methodologies, including extensions to other tools, to address PM-related issues in the Linux PM ecosystem based on the detailed bug patterns identified in this work. More importantly, 
we hope that by sharing our study and releasing the characterized dataset, our efforts could facilitate follow-up research in the community and contribute to the development of effective PM bug detection tools and the enhancement of  PM-based  systems in general}.

%% file: acknowledge.tex
\section{Acknowledgements}
\label{sec:acknowledge}

{We thank the TOS reviewers 
 and editors for their insightful and constructive feedback.} 
Also, we would like to thank researchers from Western Digital including Adam Manzanares,  Filip Blagojevic, Qing Li, and Cyril Guyot for valuable discussions on  PM technologies.
In addition, we thank Prakhar Bansal and Joshua Kalyanapu for running experiments on PM bug detectors.
This work was supported in part by NSF under grants CNS-1566554, CNS-1855565,  CNS-1943204, and a gift from Western Digital/IDEMA.
Any opinions, findings, and conclusions expressed in this material are those of the authors and do not necessarily reflect the views of the sponsor.